\documentclass[aps,amsmath,amssymb,superscriptaddress,nofootinbib]{revtex4-2}

\usepackage{graphicx}
\usepackage{color,soul}
\usepackage{morefloats}
\usepackage[percent]{overpic}
\usepackage{dcolumn}
\usepackage{multirow}

\definecolor{blue2}{rgb}{0, 0.4470, 0.7410}
\definecolor{red2}{rgb}{0.8500, 0.1250, 0.0480} 
\definecolor{orange}{rgb}{0.8500, 0.3250, 0.0980} 
\definecolor{yellow}{rgb}{0.9290, 0.6940, 0.1250}
\definecolor{purple}{rgb}{0.4940, 0.1840, 0.5560}
\definecolor{green}{rgb}{0.0, 0.5, 0.05}
\definecolor{ltblue}{rgb}{0.3010, 0.7450, 0.9330}
\definecolor{dkred}{rgb}{0.6350, 0.0780, 0.1840}
\definecolor{gray}{rgb}{0.22, 0.22, 0.3}

\newcommand{\dg}{^{\circ}}
\newcommand{\al}{\alpha}

\newcolumntype{M}[1]{>{\centering\arraybackslash}m{#1}}

\def\drawline#1#2{\raise 2.0pt\vbox{\hrule width #1pt height #2pt}}

\DeclareMathOperator*{\argmax}{arg\,max} 

\begin{document}
\title{Resolvent analysis of an airfoil laminar separation bubble at $Re = 500,000$}


\author{Chi-An Yeh}
\email[]{cayeh@seas.ucla.edu}
\affiliation{Mechanical and Aerospace Engineering, University of California, Los Angeles, CA 90095, USA}

\author{Stuart I. Benton}
\email[]{stuart.benton.1@us.af.mil}
\affiliation{Air Force Research Laboratory, Wright-Patterson Air Force Base, OH 45433, USA}

\author{Kunihiko Taira}
\email[]{ktaira@seas.ucla.edu}
\affiliation{Mechanical and Aerospace Engineering, University of California, Los Angeles, CA 90095, USA}

\author{Daniel J. Garmann}
\email[]{daniel.garmann@us.af.mil}
\affiliation{Air Force Research Laboratory, Wright-Patterson Air Force Base, OH 45433, USA}

\date{\today}

\begin{abstract}
We perform resolvent analysis to examine the perturbation dynamics over the laminar separation bubble (LSB) that forms near the leading edge of a NACA 0012 airfoil at a chord-based Reynolds number of 500,000 and an angle of attack of $8$ degrees. While we focus on the LSB residing over $6\%$ of the chord length, the resolvent operator is constructed about the global mean flow over the airfoil, avoiding numerical issues arising from domain truncation.  Moreover, randomized SVD is adopted in the present analysis to relieve the computational cost associated with the high-Re global base flow. To examine the local physics over the LSB, we consider the use of exponential discounting to limit the time horizon that allows for the instability to develop with respect to the base flow.  With discounting, the gain distribution over frequency accurately captures the spectral content over the LSB obtained from flow simulation.  The peak-gain frequency also agrees with previous flow control results on suppressing dynamic stall over a pitching airfoil.  According to the gain distribution and the modal structures, we conclude that the dominant energy-amplification mechanism is the Kelvin--Helmholtz instability.  In addition to discounting, we also examine the use of spatial windows for both the forcing and response.  From the response-windowed analysis, we find that the LSB serves the main role of energy amplifier, with the amplification saturating at the reattachment point.  The input window imposes the constraint of surface forcing, and the results show that the optimal actuator location is slightly upstream of the separation point.  The surface-forcing mode also suggest the optimal momentum forcing in the surface-tangent direction, with strong uni-directionality that is ideal for synthetic-jet-type actuators.  This study demonstrates the strength of randomized resolvent analysis in tackling high-Reynolds-number base flows and calls attention to the care needed for base-flow instabilities.  The physical insights provided by the resolvent analysis can also support flow control studies that target the LSB for suppressing flow separation or dynamic stall.
\end{abstract}

\maketitle


\section{Introduction}

The physics of an initially laminar boundary layer separating from a surface, transitioning to a turbulent state, and reattaching to the surface have been studied since at least the late 1920s \citep{Fage:PM1929, Tani:PAS1964}.  The presence of these so-called laminar separation bubbles (LSBs) is highly dependent on local Reynolds number, pressure gradient, surface curvature, and free-stream disturbance levels.  Despite typically being considered a low-Reynolds-number feature, for moderate to high incidence and leading-edge curvature, the presence of a LSB has been confirmed for airfoil chord-based Reynolds numbers approaching $10^7$ \citep{McCullough:NACA1955, Tani:PAS1964}.  Early classification of airfoil stall behaviors \citep{Jones:JRAS1934, McCullough:NACA1951} linked post-stall lift-curve shape to the presence and dynamics of the LSB during the stall event.  When LSBs exist at high Reynolds numbers on airfoils, their size can shrink to less than $1\%$ of the airfoil chord, making their detection and analysis through standard experimental techniques difficult \citep{Raffel:MST2006, Kruse:AIAA2008}.

Later, studies have highlighted the natural behavior of LSBs strongly amplifying incoming disturbances within a selective frequency range due to inviscid instability within the separated laminar shear layer \citep{Gaster:Thesis1963, Alam:JFM2000}.  This leads to a strong sensitivity to free-stream turbulence levels such that the stream-wise length of the LSB can be a function of both pressure gradient and turbulence intensity \citep{Dovgal:PAS1994, Ol:AIAA2005, Hosseinverdi:JFM2019}.  These unique aspects of LSBs have motivated numerous experiments in controlled environments, linear stability investigations (both local and global), and direct numerical simulations \citep{Haggmark:PTRSL2000, Theofilis:PTRSLA2000, Alizard:PoF2009, Maucher:AIAAJ2000, Marxen:JFM2010}.  On a related effort, unsteady forcing as a means of flow control which specifically targets shear-layer instabilities has been well studied \citep{Greenblatt:PAS2000}. In this approach, laminar separations within the airfoil boundary layer naturally amplify incoming disturbances within a specific frequency range. This generates a series of spanwise vortical structures which allow for momentum transfer from the free stream. A promising aspect of this approach is that the natural amplification by the instability allows for amplitude growth of multiple orders of magnitude such that actuator amplitude can be very minimal yet effective.

A recent campaign of research has applied wall-resolved large-eddy simulation to the study of flows over airfoils at low-Mach number and chord Reynolds numbers in the range of $Re_{L_c}=0.2-1.0\times10^6$. The airfoil is pitched about its quarter-chord axis to observe the series of nonlinear events leading to the onset of the dynamic stall process once the airfoil has pitched above its static stall angle. These computations \citep{VisbalGarmann:AIAAJ2018, Benton:JFM2019} highlighted the role of a small LSB which initiates leading-edge breakdown and the subsequent development of the dynamic stall vortex (DSV). At higher Reynolds numbers, where turbulent separation is able to progress to the airfoil leading-edge region, the breakdown of the LSB still results in a separate vortex structure \citep{Benton:JFM2019}.  Parallel efforts by the same group \citep{VisbalBenton:AIAAJ2018, BentonVisbal:PRF2018, Benton:AIAAJ2019} studied the ability for low-amplitude forcing upstream of the leading-edge LSB, in the frequency range amplified by the LSB, to alter the state of the airfoil boundary layer and significantly affect the development of dynamic stall.  This type of forcing can effectively eliminate the development of the DSV for sinusoidal pitching motions.  A further study by \citet{BentonVisbal:AIAA2018} utilizes a surface heat-flux oscillation based on the possibility of implementing a so-called `thermophone' \citep{Arnold:PRB1917, Shinoda:Nature1999, Tian:ACSNANO2011, Bin:JAP2015, Yeh:AIAA2015} for flow control.  It demonstrated that effective control could be maintained with an order-of-magnitude in energy savings, given the proper placement and tracking of the shear-layer frequency.  The control mechanism for this thermal-based actuator has been investigated in \citet{Yeh:JFM2017}, relating the thermal input to the vorticity generation near the actuator.  

These efforts suggest that the knowledge of the energy-amplification characteristics and the physical mechanism that supports the amplification over the LSBs is critical for guiding the design of active flow control.  Moreover, a challenge with developing optimal flow control techniques is the large number of flow control parameters, including actuator geometry, waveform characteristics, and actuation amplitudes.  Extensive parametric studies based on experiments or high-fidelity numerical simulations are costly.  The practical constraints on the actuator types, sizes, and equipments also need to be taken into account in the investigation.  

In the present effort, we leverage resolvent analysis to conduct preliminary investigations into the unsteady response of a given flow field in order to reduce the parameter space to search for the optimal control setup.  Among the modal analysis techniques \citep{Taira:AIAAJ2017, Taira:AIAAJ2019}, resolvent analysis is an attractive tool for this investigation since it is concerned with the energy amplification in an input--output process.  It allows for the characterization of the frequency response of a base flow \citep{Jovanovic:JFM2005, Trefethen:Science1993}.  The corresponding modal shapes provide physical insights on the mechanism of how the perturbations are amplified or attenuated.  This approach is closely related to the pseudospectral analysis by \citet{Trefethen:Science1993}, which provides physics-based explanation on subcritical transition in turbulent flows.  Moreover, \citet{McKeonSharma:JFM2010} have extended resolvent analysis to turbulent flows by considering the finite-amplitude nonlinear interaction terms as a self-sustained internal forcing.

The present study aims to investigate the perturbation dynamics of the LSB that forms in a high-Reynolds-number flow near the leading edge of a canonical airfoil.  Since the high-Reynolds-number flow considered in this study is turbulent, we adopt the perspective of \citet{McKeonSharma:JFM2010} to conduct the resolvent analysis of the LSB.  Moreover, we adopt discounted resolvent analysis to address the input-output relation for the unstable linear operator \citep{Jovanovic:Thesis2004, Yeh:JFM2019}.  The use of spatial window is also considered to uncover the mechanism of energy-amplification over the LSB \citep{Jeun:PoF2016, Schmidt:JFM2018} and to shed light on the placement of an actuator.  We will find that the energy amplification saturates after the flow reattaches, which suggests that the LSB is indeed the main energy amplifier.  We will also show that the optimal actuator location is upstream of the separation point, with the most effective momentum-based forcing aligning with the direction of the local surface tangent.

\section{Approach}

\subsection{Problem setup}

We analyze the perturbation dynamics over the laminar separation bubble that forms over a NACA 0012 airfoil at an angle of attack $\alpha = 8^{\circ}$.  The flow over the airfoil is analyzed at a chord-based Reynolds number of $Re_{L_c} \equiv v_\infty L_c/\nu_\infty=500,000$ and a free-stream Mach number of $M_\infty \equiv v_\infty/a_\infty = 0.1$, where $L_c$ is the chord length of the airfoil, and $v_\infty$, $\nu_\infty$ and $a_\infty$ are respectively the free-stream velocity, kinematic viscosity and sonic speed.  
To perform this study, we leverage implicit large-eddy simulation (ILES), linear stability analysis, and resolvent analysis, as described below.  Throughout this study, the Cartesian coordinate system used where the $x$, $y$, and $z$ directions are aligned with the streamwise, transverse, and spanwise directions, respectively.

\subsection{Large-eddy simulation}

The base flow used in the present modal analyses is the time- and span-average of the ILES computation presented in \citet{BentonVisbal:AIAA2018}, where the $Re_{L_c} = 500,000$ flow over the airfoil is examined with an ILES based on the extensively validated high-order Navier-Stokes equation solver \textit{FDL3DI} \citep{VisbalGaitonde:AIAAJ1999, Gaitonde:AFRL1998}.  In this code, a sixth-order compact finite-difference scheme \citep{Lele:JCP1992} is employed to discretize the governing equations, along with a high-order lowpass spatial filtering \citep{VisbalGaitonde:AIAAJ1999, Gaitonde:IJNME1999} to eliminate spurious components.  The filtering operation is applied to the conserved variables along each transformed coordinate direction once after each time step or sub-iteration.  Time-marching is accomplished through the second-order, iterative, implicit, approximately-factored Beam and Warming method \citep{Beam:AIAAJ1978}.  

In order to perform an ILES, the above numerical methods are applied to the original \textit{unfiltered} Navier--Stokes equations, and are used without change in laminar, transitional, or fully turbulent regions of the flow.  For transitional and turbulent regions, these high-fidelity spatial algorithmic components provide an effective ILES approach in lieu of traditional sub-grid stress models \citep{Visbal:JFE2002, Rizzetta:IJNMF2003, Garmann:IJNMF2013}.  In regions of laminar and early-stage transition, this methodology is effectively a direct numerical simulation of the Navier--Stokes equations. The transition over the LSB occurred naturally in the simulation without explicit forcing.

\subsection{Quasi-parallel linear stability analysis}

\begin{figure}
    \centering
    \includegraphics[width=1.0\textwidth, trim=0.162in 0.7in 0.162in 0, clip]{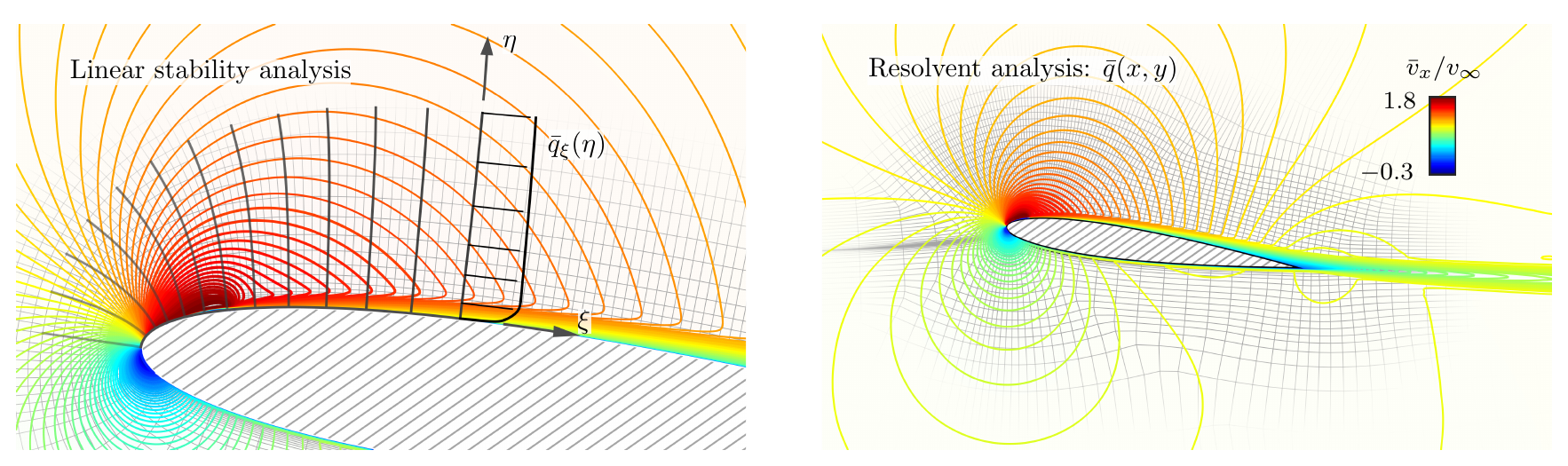}
    \caption{\label{fig:Demo_LSA_vs_Resolvent} The grid setups for the linear stability analysis (left) and bi-global resolvent analysis (right). The mean flow obtained on the LES mesh is interpolated onto a separate grid to conduct the resolvent analysis.  Every 8 grid lines are shown.  Contour lines are shown for the time- and spanwise-averaged streamwise velocity.}
\end{figure}

We perform the quasi-parallel linear stability analysis on the time-averaged boundary layer profiles over the LSB using the linearized compressible Navier--Stokes equations.  On a body-fitted coordinate system with its origin located at the stagnation point, this quasi-parallel stability analysis considers infinitesimal perturbations $q_\xi'(\eta)$ about the wall-normal boundary layer profiles $\bar{q}_\xi(\eta)$ at successive wall-tangent stations $\xi$, as shown in FIG. \ref{fig:Demo_LSA_vs_Resolvent} (left).  By plugging in a real frequency $\omega$ and a real spanwise wavenumber $k_z$, the spatial stability analysis is performed by formulating 
\begin{equation}\label{eq:LASTRAC_LST}
	\left[-i \omega - \boldsymbol{\mathcal{L}}_{\bar{q}_\xi}(k_x, k_z) \right]\hat{q}_\xi =  0,
\end{equation}
as an eigenvalue problem with respect to the complex streamwise wavenumber $k_x = k_{x,r} + i k_{x,i}$.  Here, $\boldsymbol{\mathcal{L}}_{\bar{q}_\xi} \equiv k_x^2 \boldsymbol{\mathcal{L}}_2(k_z) - k_x \boldsymbol{\mathcal{L}}_1(k_z) - \boldsymbol{\mathcal{L}}_0(k_z)$ is the linearized Navier--Stokes operator about the boundary layer profile $\bar{q}_\xi(\eta)$ with the boundary conditions of $q'_\xi = 0$ at $\xi = 0$ and $\xi \rightarrow \infty$, and $\hat{q}_\xi(\eta)$ is the Fourier mode representation of $q'_\xi(\eta) = \hat{q}_\xi(\eta) \exp \left[i(k_zz - k_xx - \omega t)\right]$.  
For each combination of streamwise station and frequency, the most unstable eigenvalue (with the highest $k_{x, i}$) is extracted and saved. For each frequency, the local spatial amplification rate $k_{x,i}(\xi)$ is integrated over $\xi \in [\xi_0, \xi]$ such that the total $\xi$-wise amplification (effective gain) of an incoming disturbance $a_0$ can be expressed as
\begin{equation}
	a(\omega, \xi) / a_0 = e^{\int_{\xi_0}^{\xi}k_{x,i}\left(\omega, \xi\right){\rm d}\xi}.
\label{eq:ampLST}
\end{equation}
where $\xi_0$ is the initial point of integration (the first station to return an unstable eigenvalue).  The results obtained from the quasi-parallel linear stability analysis will be used to identify relevant frequency range of LSB dynamics.  We will sweep over this frequency range in fine increment in the computationally intensive global resolvent analysis, and examine what can be further revealed by the global analysis from the quasi-parallel linear stability analysis.

\subsection{Resolvent analysis}
\subsubsection{Formulation}

We conduct the global resolvent analysis about the mean turbulent flow.  
The flow variable is Reynolds decomposed into the sum of the time- and spanwise-averaged base state $\bar{\boldsymbol{q}}(x, y) \equiv [\bar{\rho}, \bar{v}_x, \bar{v}_y, \bar{v}_z, \bar{p}]$ and the statistically stationary fluctuating component $\boldsymbol{q}(x, y, z, t) \equiv \left[\rho', v_x', v_y', v_z', T'\right]$.  
With the Reynolds decomposition, we express the compressible Navier--Stokes equations as an input-output system \citep{Trefethen:Science1993, Jovanovic:JFM2005, McKeonSharma:JFM2010, Jeun:PoF2016, Schmidt:JFM2018, Kojima:JFM2020}
\begin{subequations}
\label{eq:io_time}
\begin{align}
    \partial_t \boldsymbol{q} &= \boldsymbol{\mathcal{A}}_{\bar{\boldsymbol{q}}} \boldsymbol{q}+ \boldsymbol{B}\boldsymbol{f},\\
    \boldsymbol{y} &= \boldsymbol{C}\boldsymbol{q},
\end{align}
\end{subequations}
where $\boldsymbol{\mathcal{A}}_{\bar{\boldsymbol{q}}}$ is the linearized Navier--Stokes operator constructed about the two-dimensional mean flow $\bar{\boldsymbol{q}}(x,y)$.  When a time-averaged flow is used in resolvent analysis, the finite-amplitude nonlinear terms with respect to $\boldsymbol{q}$ are collected in $\boldsymbol{u}$, which can be viewed as a sustained internal forcing input within the natural feedback system about the mean state \citep{FarrellIoannou:PRL1994,Schmid:ARFM2007,McKeonSharma:JFM2010}.  In the above input-output formulation, the matrices $\boldsymbol{B}$ and $\boldsymbol{C}$ specifies the input and output of interest, which will be discussed later in section \ref{sec:windowing}.

For the variables in the input-output system (\ref{eq:io_time}), we perform Laplace transform in time and Fourier transform in the homogeneous $z$-direction to yield
\begin{equation}
	\hat{\boldsymbol{q}}(x, y)
		= \int_0^\infty \int_{-\infty}^{\infty} \boldsymbol{q}(\boldsymbol{x}, t) e^{-ik_z z - s t} {\rm d}z {\rm d}t,
\end{equation}
where $s$ is the Laplace variable and the same transformation applies to $\boldsymbol{y}$ and $\boldsymbol{f}$.  With this Laplace--Fourier transform, the asymptotic  output under the sustained harmonic input can be found in the frequency domain as
\begin{equation}
	\hat{\boldsymbol{y}}
	= \boldsymbol{\mathcal{H}}_{\bar{\boldsymbol{q}}}(s, k_z)
	\hat{\boldsymbol{f}},~~\text{where}~~
	\boldsymbol{\mathcal{H}}_{\bar{\boldsymbol{q}}}(s, k_z) \equiv \boldsymbol{C} \left[ s\boldsymbol{I} - \boldsymbol{\mathcal{A}}_{\bar{\boldsymbol{q}}}(k_z) \right]^{-1} \boldsymbol{B}.
	\label{eq:ns-resol}
\end{equation}
The operator $\boldsymbol{\mathcal{H}}_{\bar{\boldsymbol{q}}}(s, k_z)$ is referred to as the resolvent, which serves as a transfer function that amplifies (or attenuates) the harmonic forcing input $\hat{\boldsymbol{f}}$ and maps it to the response $\hat{\boldsymbol{y}}$ at a given combination of $(s, k_z)$.

Resolvent analysis identifies the dominant directions along which $\hat{\boldsymbol{f}}$ can be most amplified through $\boldsymbol{\mathcal{H}}_{\bar{\boldsymbol{q}}}(s, k_z)$ to form the corresponding responses in $\hat{\boldsymbol{y}}$.  
These directions can be determined through singular value decomposition (SVD) of
\begin{equation}
    \boldsymbol{\mathcal{H}}_{\bar{\boldsymbol{q}}}(s, k_z) = \boldsymbol{Y}\boldsymbol{\Sigma}\boldsymbol{F}^*,
    \label{eq:svd}
\end{equation}
where $\boldsymbol{F}^*$ denotes the Hermitian transpose of $\boldsymbol{F}$.  
Resolvent analysis interprets the singular vectors $\boldsymbol{Y} = [\hat{\boldsymbol{y}}_1, \hat{\boldsymbol{y}}_2, \dots, \hat{\boldsymbol{y}}_m]$ and $\boldsymbol{F} = [\hat{\boldsymbol{f}}_1, \hat{\boldsymbol{f}}_2, \dots, \hat{\boldsymbol{f}}_m]$ respectively as the response modes and forcing modes, with the magnitude-ranked singular values $\boldsymbol{\Sigma} = {\text{diag}}(\sigma_1, \sigma_2, \dots, \sigma_m) $ being the amplifications (gains) for the corresponding forcing--response pair.   In this study, we examine the resolvent gains and modes in detail to study the perturbation dynamics over the laminar separation bubble.

\subsubsection{Stability of $\boldsymbol{\mathcal{A}}_{\bar{\boldsymbol{q}}}$ and discounting}

\begin{figure}
\centering
    \includegraphics[width=1.0\textwidth, trim=0.162in 0.6in 0.162in 0, clip]{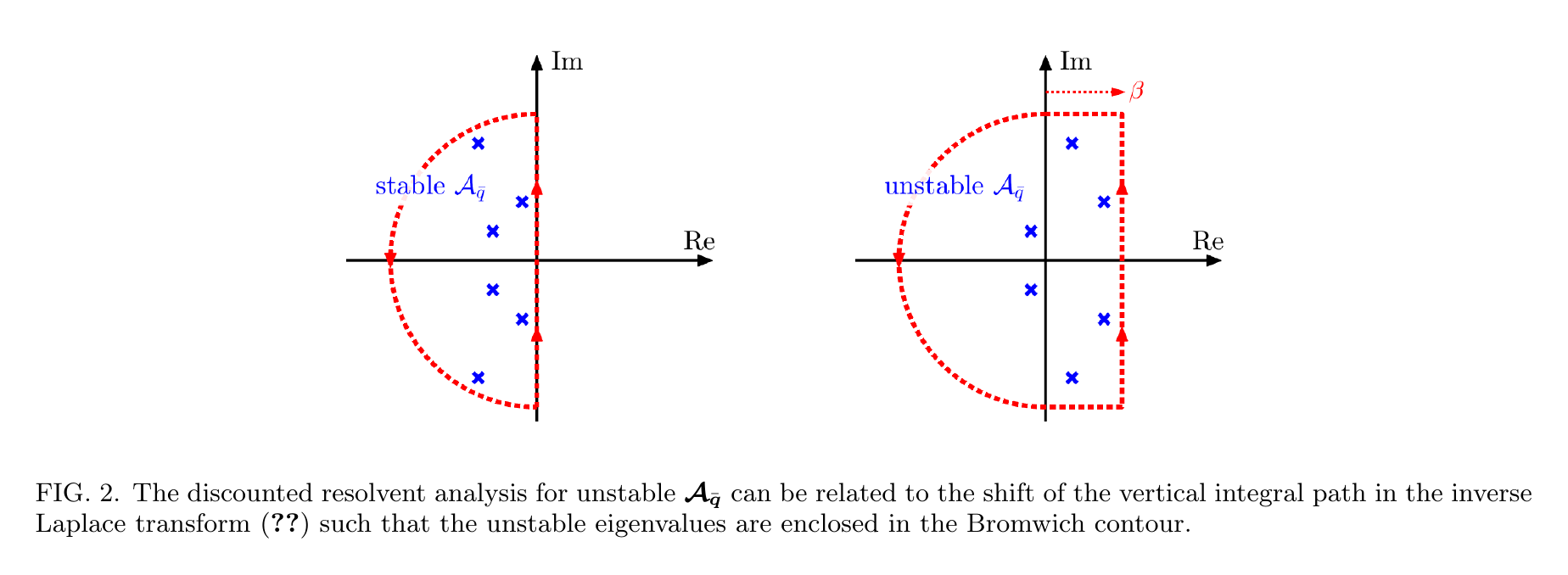}
	\caption{\label{fig:disc_demo} The discounted resolvent analysis for unstable $\boldsymbol{\mathcal{A}}_{\bar{\boldsymbol{q}}}$ can be related to the shift of the vertical integral path in the inverse Laplace transform (\ref{eq:inv_Laplace}) such that the unstable eigenvalues are enclosed in the Bromwich contour.}
\end{figure}

We can bring the output $\hat{\boldsymbol{y}}$ in equation (\ref{eq:ns-resol}) back to the time domain by performing the inverse Laplace transform using the Bromwich integral 
\begin{equation}
    \boldsymbol{y}(t) =  \frac{1}{2\pi i} \oint_{{\scriptsize\reflectbox{D}} \rightarrow \infty}  \boldsymbol{\mathcal{H}}_{\bar{\boldsymbol{q}}}(s) \hat{\boldsymbol{f}}(s) {\rm d}s = \frac{1}{2\pi i} \int_{\beta - i\infty}^{\beta + i\infty} \boldsymbol{C} \left[ s\boldsymbol{I} - \boldsymbol{\mathcal{A}}_{\bar{\boldsymbol{q}}} \right]^{-1} \boldsymbol{B} \hat{\boldsymbol{f}}(s) e^{st} {\rm d}s,
    \label{eq:inv_Laplace}
\end{equation}
where we have left out the $k_z$-dependence for brevity.  To satisfy causality, the real-valued parameter $\beta$ needs to be chosen such that all poles of the transfer function $\boldsymbol{\mathcal{H}}_{\bar{\boldsymbol{q}}}(s)$ reside on the left side of the line of integration $s(\beta, \omega) = \left\lbrace\beta + i\omega~|~\omega \in (-\infty, \infty) \right\rbrace $, allowing for the \reflectbox{D}-shaped Bromwich contour to enclose all the poles.  This is demonstrated in FIG. \ref{fig:disc_demo}.  It is important to include the right-most poles in the Bromwich contour, since in general they are the most dominant ones in determining the dynamical behavior of the system response.  According to equation (\ref{eq:inv_Laplace}), these poles are the eigenvalues of $\boldsymbol{\mathcal{A}}_{\bar{\boldsymbol{q}}}$, suggesting that $\beta$ needs to be greater than the dominant modal growth rate of $\boldsymbol{\mathcal{A}}_{\bar{\boldsymbol{q}}}$.  

Resolvent analysis for fluid-flow problems is usually conducted along this line of integration, which is equivalent to sweeping through the frequency (imaginary) component of the Laplace variable, $s = \beta + i\omega$.  
Commonly, $\beta$ is chosen to be zero, and the Laplace transform in (\ref{eq:inv_Laplace}) degrades to the Fourier transform \citep{McKeonSharma:JFM2010,Luhar:JFM2015,ThomareisPapadakis:PRF2017}.  Fourier transform is a natural choice for a statistically stationary flow and is appropriate for a stable $\boldsymbol{\mathcal{A}}_{\bar{\boldsymbol{q}}}$ as all of its eigenvalues lie on the left plane of the imaginary axis.  However, statistical stationarity does not guarantee a stable $\boldsymbol{\mathcal{A}}_{\bar{\boldsymbol{q}}}$.  If $\boldsymbol{\mathcal{A}}_{\bar{\boldsymbol{q}}}$ has unstable eigenvalues, the use of $\beta = 0$ does not recover the output $\boldsymbol{y}(t)$, as suggested by equation (\ref{eq:inv_Laplace}).  

Therefore, when performing resolvent analysis for an unstable system along a path parallel to the frequency axis, we consider a shift of this path to the right of the most unstable eigenvalue by choosing $\beta > \max\left(\Re(\lambda)\right)$, where $\lambda$ are the eigenvalues of $\boldsymbol{\mathcal{A}}_{\bar{\boldsymbol{q}}}$.  The effect of this shift is equivalent to the exponential discounting approach introduced by \citet{Jovanovic:Thesis2004}, where a temporal damping $e^{-\beta t}$ is applied to the variables as $\left[\boldsymbol{y}_\beta,\boldsymbol{q}_\beta,\boldsymbol{f}_\beta\right] = e^{-\beta t}\left[\boldsymbol{y},\boldsymbol{q},\boldsymbol{f}\right]$.  Substituting the discounted variables $\left[\boldsymbol{y}_\beta,\boldsymbol{q}_\beta,\boldsymbol{f}_\beta\right]$ into equation (\ref{eq:io_time}) and performing the Laplace transform, the discounted resolvent operator can be found as
\begin{equation}
	\boldsymbol{\mathcal{H}}_{\bar{\boldsymbol{q}}}(\beta, s, k_z) \equiv \boldsymbol{C} \left[ (s + \beta)\boldsymbol{I} - \boldsymbol{\mathcal{A}}_{\bar{\boldsymbol{q}}}(k_z) \right]^{-1} \boldsymbol{B},
	\label{eq:resol_disc}
\end{equation}
and the shifted path in FIG. \ref{eq:inv_Laplace} can be recovered by considering $s = i\omega$.  The discounted resolvent operator (\ref{eq:resol_disc}) can be equivalently expressed as $\boldsymbol{\mathcal{H}}_{\bar{\boldsymbol{q}}} \equiv \boldsymbol{C} \left[ i\omega \boldsymbol{I} -\left( \boldsymbol{\mathcal{A}}_{\bar{\boldsymbol{q}}} - \beta \boldsymbol{I} \right)\right]^{-1} \boldsymbol{B}$, representing the transfer function for the harmonic input--output process through the stable operator $\left( \boldsymbol{\mathcal{A}}_{\bar{\boldsymbol{q}}} - \beta \boldsymbol{I} \right)$.This discounting approach can be viewed as a finite-time horizon input-output analysis, since there exists a finite time $t$ such that the monotonically increasing gain for the unstable system at $t$ is equal to the long-time amplification from a corresponding discounted system \citep{Jovanovic:Thesis2004}.  We also note that discounted forcing implies that the forcing amplitude grows faster than the instabilities within the unstable system.  This ensures that the amplification of forcing overtakes the growth of the instabilities, which is an important perspective in examining the frequency response to external forcing for flow control.  In our previous effort \citep{Yeh:JFM2019}, we also found that the effect of a higher discounting parameter $\beta$ is similar to that of a shorter time-horizon over which the exponentially increasing gain is evaluated, and the streamwise extent of convective response structure reduces with the higher $\beta$.  This finding further motivates the use of the discounted analysis to reveal the local physics of the laminar-separation bubble from a global resolvent operator by choosing an appropriate range for the discounting parameter $\beta$.

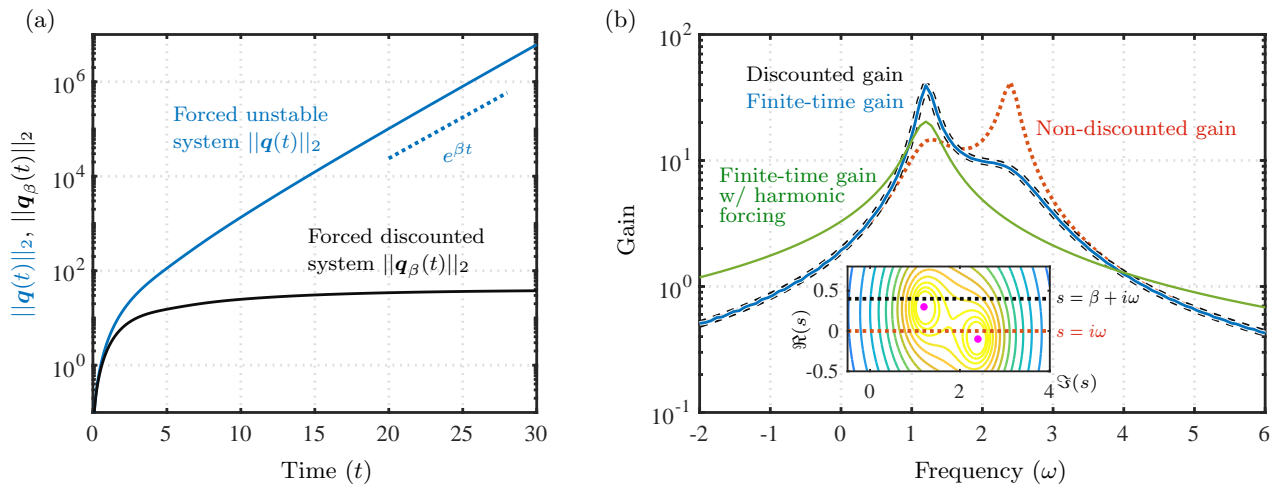
\begin{figure}
\centering
    \begin{overpic}[width=1.0\textwidth]{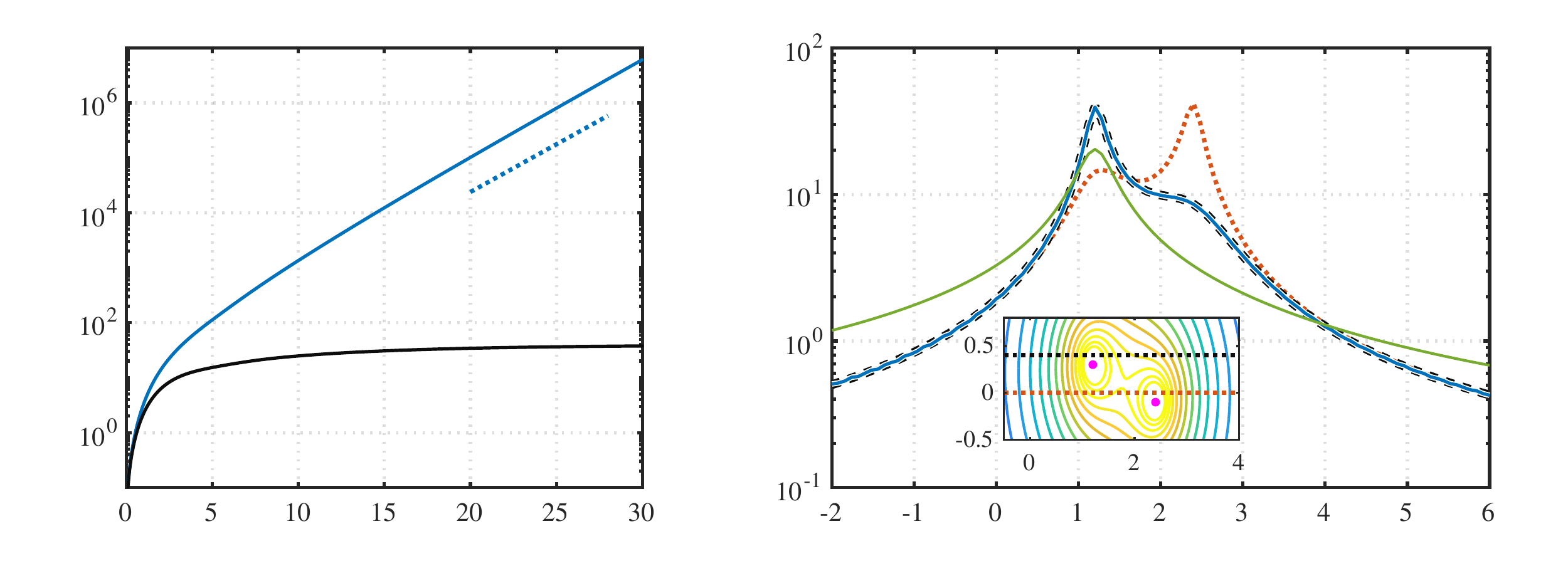}
		\put(03, 35){(a)}
		\put(22, 1.5){Time ($t$)}
		\put(02, 13.4){\rotatebox{90}{{\color{blue2}$||\boldsymbol{q}(t)||_2$}, $||\boldsymbol{q}_\beta(t)||_2$ }}
		\put(14, 28){\footnotesize\color{blue2} Forced unstable}
		\put(14, 26){\footnotesize\color{blue2} system $||\boldsymbol{q}(t)||_2$}
		\put(34, 25){\footnotesize\color{blue2} $e^{\beta t}$}
		\put(24, 19){\footnotesize Forced discounted}
		\put(24, 17){\footnotesize system $||\boldsymbol{q}_\beta(t)||_2$}
		
		\put(46, 35){(b)}
		\put(69, 1.5){Frequency ($\omega$)}
		\put(47, 18){\rotatebox{90}{Gain}}
		\put(78, 27){\footnotesize\color{red2} Non-discounted gain}
		\put(56.5, 31){\footnotesize Discounted gain}
		\put(56.5, 29){\footnotesize\color{blue2} Finite-time gain}
		\put(54.5, 23.5){\footnotesize\color{green} Finite-time gain}
		\put(54.5, 22.0){\footnotesize\color{green} w/ harmonic}
		\put(54.5, 20.5){\footnotesize\color{green} forcing}
		
		\put(79.5,  8.2){\scriptsize$\Im(s)$}
		\put(59.9, 11.0){\rotatebox{90}{\scriptsize$\Re(s)$}}
		\put(79.5, 14.5){\scalebox{.85}[1.0]{\scriptsize $s = \beta+i\omega$}}
		\put(79.5, 12.0){\scalebox{.85}[1.0]{\scriptsize\color{red2} $s = i\omega$}}
	\end{overpic}
	\caption{\label{fig:disc_2x2} Discounted resolvent analysis of a simple system: (a) Finite-time energy amplification for the forced unstable system and discounted stable system with forcing frequency $\omega = 1.2$; (b) Comparison of the finite-time gain from the unstable system (evaluated at $t = 30$ and rescaled as $e^{-\beta t}|| \boldsymbol{q} ||_2$ ) to the gains obtained from discounted and non-discounted resolvent analyses.  Finite-time gain of the harmonically forced unstable system   (evaluated at $t = 30$ and rescaled as $e^{-\Re(\lambda_1) t}|| \boldsymbol{q} ||_2$) also shows agreement with the gain profile obtained from the discounted resolvent analysis.  Inserted in (b) are the eigenvalues (magenta dots) and pseudospectrum (contour levels) of $\boldsymbol{\mathcal{A}}_{\text{u}}$.  The non-discounted analysis is performed over the path of $s = i\omega$ and the discounted analysis is performed over $s = \beta+i\omega$.}
\end{figure}

Let us use a simple example to demonstrate how the discounting approach captures the energy amplification in a forced unstable system.   We consider the forced problem of $\partial_t \boldsymbol{q} = \boldsymbol{\mathcal{A}}_{\text{u}}\boldsymbol{q} + \boldsymbol{f}$, where
\begin{equation*}
	\boldsymbol{\mathcal{A}}_{\text{u}} = 
	\begin{bmatrix} 
		0.3+1.2i & 5 \\
		0 & -0.1+2.4i
\end{bmatrix}
	\label{eq:A2by2}
\end{equation*}
has an unstable pole at $\lambda_1 = 0.3+1.2i$.  For this system we choose a discounting parameter $\beta = 0.4 > \Re(\lambda_1)$ and find the discounted input-output system $\partial_t \boldsymbol{q}_\beta = \boldsymbol{\mathcal{A}}_{\text{d}}\boldsymbol{q}_\beta + \boldsymbol{f}_\beta$, where $\boldsymbol{\mathcal{A}}_{\text{d}} = \boldsymbol{\mathcal{A}}_{\text{u}} - \beta\boldsymbol{I}$ is stable.  Starting with zero initial condition, we force both the unstable and discounted systems at a single frequency $\omega = 1.2$ and show the energy growths of $\boldsymbol{q}$ and $\boldsymbol{q}_\beta$ in FIG. \ref{fig:disc_2x2}a.  The asymptotic value of $||\boldsymbol{q}_\beta||_2$ reveals the gain at the forcing frequency for the stable discounted system \citep{SchmidHenningson:2001}.  For the original unstable system, we can also choose a finite time horizon over which we examine the effect of forcing frequency on the exponentially growing $||\boldsymbol{q}||_2$.  This finite-time gain profile, determined with $||\boldsymbol{q}(t)||_2$ at $t = 30$, is accurately captured by the gain from the discounted resolvent analysis performed over the path of $s = \beta + i\omega$, as shown in FIG. \ref{fig:disc_2x2}b.  However, the gain profile obtained from the non-discounted resolvent analysis is completely off from the finite-time gain profile obtained from the simulation of the forced unstable system.  In addition to forcing the unstable system with growing amplitude, we also consider the use of harmonic forcing (constant forcing amplitude in time) to the unstable system.  The trend of the finite-time amplification for the harmonically forced unstable system with respect to the forcing frequency is captured by the discounted analysis, showing the most amplified forcing frequency at $\omega = 1.2$.  These insights given by the discounted resolvent analysis are crucial from the perspectives of flow control where the trend in gain profiles is relied on.

Unlike an equilibrium base flow\footnote{For unsteady flows, the equilibrium base states, also referred to as the steady solutions or the fixed points, can be found through selective frequency damping \citep{Akervik:PF06} or Newton-based solvers \citep{Kelley:2004}. Their stabilities can usually be inferred by how the flows naturally behave without any external forcing \citep{Zhang:PoF2016,Sun:JFM2017}.}, inferring the stability of a mean flow (or, more precisely, the stability of $\boldsymbol{\mathcal{A}}_{\bar{\boldsymbol{q}}}$ constructed about the mean flow $\bar{\boldsymbol{q}}$) via physical intuition is not always straightforward.  Thus, a companion stability analysis of $\boldsymbol{\mathcal{A}}_{\bar{\boldsymbol{q}}}$ is required to identify the most unstable eigenvalue, with which the minimal discounting parameter $\beta$ can be determined.  In the present study,  we perform the stability analysis of $\boldsymbol{\mathcal{A}}_{\bar{\boldsymbol{q}}}$ as a precursor to resolvent analysis, and the dominant global modes will also be discussed.  However, we also note that stability analysis of time-averaged flow has been questioned for its validity of providing physical interpretations, since such a base flow does not constitute a steady solution to the Navier--Stokes equations due to the omitted nonlinear terms of finite amplitude. \citep{Sipp:JFM2007,Beneddine:JFM2016,Taira:AIAAJ2017} .

\subsubsection{Windowed analysis}
\label{sec:windowing}
\begin{figure}
    \includegraphics[width=1.0\textwidth, trim=0.162in 0.7in 0.162in 0.1in, clip]{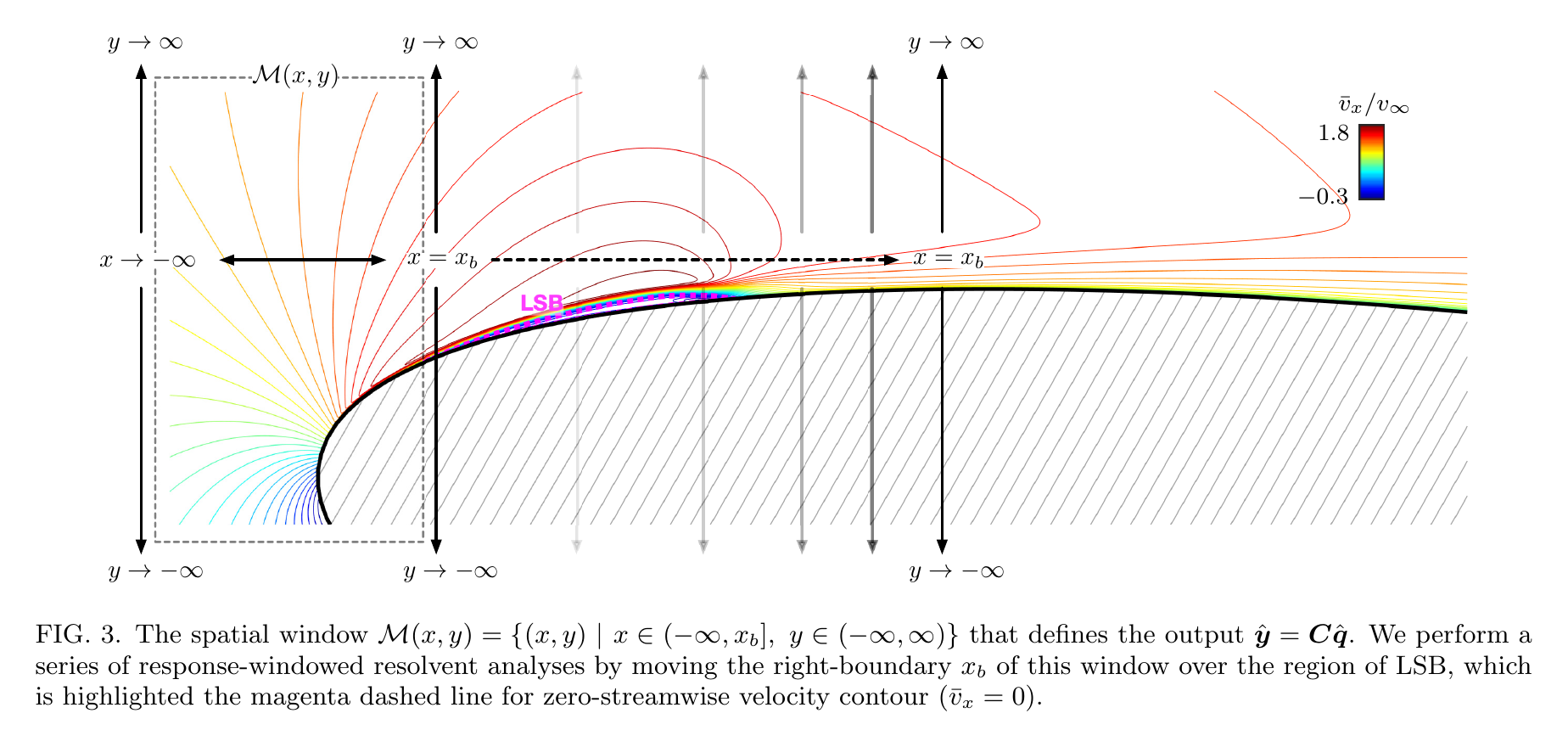}
	\caption{\label{fig:window_demo} The spatial window $\mathcal{M}(x, y) = \{(x, y)~|~x \in (-\infty, x_b],~y \in (-\infty, \infty)\}$ that defines the output $\hat{\boldsymbol{y}} = \boldsymbol{C} \hat{\boldsymbol{q}}$.  We perform a series of response-windowed resolvent analyses by moving the right-boundary $x_b$ of this window over the region of LSB, which is highlighted the magenta dashed  line for zero-streamwise velocity contour ($\bar{v}_x = 0$). }
\end{figure}

For studying the local perturbation dynamics of LSB, we also consider the use of spatial windows in the present input-output analysis \citep{Jeun:PoF2016}.  To identify the region that is associated with high energy-amplification over the LSB and its connection with the laminar-turbulent transition and reattachment \citep{Marxen:JFM2011, Yeh:JFM2019}, we consider the output spatial window $\mathcal{M}(x, y) = \{(x, y)~|~x \in (-\infty, x_b],~y \in (-\infty, \infty)\}$ that encapsulates the region of LSB.  This spatial window is implemented in the output matrix $\boldsymbol{C}$ in equation (\ref{eq:ns-resol}) by assigning unit weights to its diagonal elements inside the window and zeros otherwise \citep{Schmidt:JFM2018,Kojima:JFM2020}.  This matrix $\boldsymbol{C}$ defines the output $\hat{\boldsymbol{y}}$ as the response observed only through this spatial window, while masking out the response outside.  Meanwhile, we keep the input matrix $\boldsymbol{B} = \boldsymbol{I}$ such that there is no spatial constraint on the forcing input.  We move the right-boundary of the spatial window $x_b$ along the streamwise extent of the LSB, as depicted in FIG. \ref{fig:window_demo}, and examine how the resolvent gain changes when $x_b$ moves over the point of laminar separation, shear-layer roll-up, transition, and reattachment. 

Similar to the output-windowed analysis, we consider the use of the input windowing is implemented in  matrix $\boldsymbol{B}$.  This input-windowed analysis is motivated by the actuator placement problem encountered in active flow control.  In practical scenarios, the actuator can only be placed on the surface of aerodynamic body.  We also note that the chosen actuator can only introduce a specific form of forcing input \citep{Cattafesta:ARFM11}, such as momentum \citep{Glezer:ARFM2002,BentonVisbal:PRF2018} or thermal \citep{Little&Samimy:AIAAJ2012, BentonVisbal:AIAA2018, Yeh:AIAA2017, Yeh:JFM2017, Little:AIAAJ2019}.  Therefore, we design the input matrix $\boldsymbol{B}$ such that the forcing can only be introduced from the surface of the airfoil with specific forcing inputs.  A similar use of the input matrix has been considered by \citet{Garnaud:JFM2013} such that the forcing can only be introduced from the flow boundary.  In addition to the spatial constraint, we add forcing only to the momentum equations by prescribing zero weights for the components in continuity and energy equations \citep{Jovanovic:JFM2005}.  

\subsubsection{Numerical setup}

We interpolate the time- and spanwise-averaged turbulent flow onto a separate mesh with its resolution specifically tailored toward resolving the short-wavelength of the high-frequency structures in the vicinity of the laminar separation bubble.  This interpolated base flow and the mesh are shown in Fig. \ref{fig:Demo_LSA_vs_Resolvent} (right).  The linearized Navier--Stokes operator $\boldsymbol{\mathcal{A}}_{\bar{\boldsymbol{q}}}$ is discretized with this mesh using a finite-volume compressible flow solver {\it CharLES} \citep{Bres:AIAAJ2017, Sun:JFM2017}, incorporating the boundary conditions of $[\rho', v_x', v_y', v_z', \nabla_n p'] = 0$ over the airfoil and the far field and $\nabla_n [\rho', v_x', v_y', v_z', p'] = 0$ at the computational outlet, where $\nabla_n$ denotes the surface-normal gradient.  This mesh has approximately $0.22 \times 10^6$ grid points and the matrix-based size of the resulting resolvent operator $\boldsymbol{\mathcal{H}}_{\bar{\boldsymbol{q}}}(s, k_z)$ is approximately $10^6 \times 10^6$.  In the resolvent computation, we incorporate the energy norm of Chu \citep{Chu:Acta1965} with the cell volume size in a weighting quadrature matrix $\boldsymbol{W}$ and perform the SVD for $\boldsymbol{W} \boldsymbol{\mathcal{H}}_{\bar{\boldsymbol{q}}} \boldsymbol{W}^{-1}$ \citep{SchmidHenningson:2001,Yeh:JFM2019}.

\subsubsection{Randomized SVD}

Performing SVD for the aforementioned high-dimensional resolvent operator is computationally taxing and memory intensive.  Since we only seek a few dominant resolvent modes, we consider leveraging randomized numerical linear algebra to relieve the computational cost of performing the large-scale SVD.
Instead of directly performing the SVD on the large $\boldsymbol{\mathcal{H}}_{\bar{\boldsymbol{q}}} \in \mathbb{C}^{m \times m}$ to find the leading-mode representation \citep{McKeonSharma:JFM2010}, we carry out the SVD on a small low-rank representation of $\boldsymbol{\mathcal{H}}_{\bar{\boldsymbol{q}}}$.  The first and essential step is to obtain a {\it sketch} $\boldsymbol{S}$ of the operator $\boldsymbol{\mathcal{H}}_{\bar{\boldsymbol{q}}}$ by passing a tall and skinny test matrix $\boldsymbol{\Omega} \in \mathbb{R}^{m \times k}$ ($k \ll m$) through 
\begin{equation}
    \boldsymbol{S} = \boldsymbol{\mathcal{H}}_{\bar{\boldsymbol{q}}} \boldsymbol{\Omega}.
\end{equation}
This provides a sketch $\boldsymbol{S}$ that contains the leading actions of $\boldsymbol{\mathcal{H}}_{\bar{\boldsymbol{q}}}$ \citep{Woolfe:ACHA08, Halko:SIAMReview11, Tropp:SIAMMAA17}.  The standard choice for a test matrix $\boldsymbol{\Omega}$ is a random matrix with a normal Gaussian distribution~\cite{Martinsson:ACHA2011}.  Here, we consider the use of physics-informed $\boldsymbol{\Omega}$ by further weighing each element of a random matrix with the corresponding velocity gradient \citep{Ribeiro:PRF2019}.

As the sketch $\boldsymbol{S}$ holds the dominant influence of $\boldsymbol{\mathcal{H}}_{\bar{\boldsymbol{q}}}$, we can form an orthonormal basis $\boldsymbol{Q} \in \mathbb{C}^{m \times k}$ from  $\boldsymbol{S}$ using QR decomposition and project the full $\boldsymbol{\mathcal{H}}_{\bar{\boldsymbol{q}}}$ onto $\boldsymbol{Q}$ to derive its low-rank approximation.  In this way, it is possible to approximate $\boldsymbol{\mathcal{H}}_{\bar{\boldsymbol{q}}}$ for a rank $k \ll m$, since this approximation preserves the features of the leading modes.  With the orthonormal basis $\boldsymbol{Q}$, a low-rank approximation of $\boldsymbol{\mathcal{H}}_{\bar{\boldsymbol{q}}}$ can be found as $\boldsymbol{\mathcal{H}}_{\bar{\boldsymbol{q}}} \approx \boldsymbol{Q} \boldsymbol{G}$, where $\boldsymbol{G} = \boldsymbol{Q}^* \boldsymbol{\mathcal{H}}_{\bar{\boldsymbol{q}}} \in \mathbb{C}^{k\times m}$ is the projection of $\boldsymbol{\mathcal{H}}_{\bar{\boldsymbol{q}}}$ onto the reduced basis \cite{Halko:SIAMReview11}.  It is this reduced matrix $\boldsymbol{G}$ upon which we perform the SVD, i.e. $\boldsymbol{G} = \boldsymbol{U}\tilde{\boldsymbol{\Sigma}}\tilde{\boldsymbol{F}^*}$, leading to the low-rank approximation of  
\begin{equation}
	\boldsymbol{\mathcal{H}}_{\bar{\boldsymbol{q}}} \approx \boldsymbol{Q}\boldsymbol{U}\tilde{\boldsymbol{\Sigma}}\tilde{\boldsymbol{F}^*},
	\label{eq:approx}
\end{equation}
where the left singular vector is approximated as $\tilde{\boldsymbol{Y}} = \boldsymbol{Q}\boldsymbol{U}$.  This is the standard process for randomized SVD by \citet{Halko:SIAMReview11}.  In this study, we follow \citet{Ribeiro:PRF2019} to recover the gain $\tilde{\boldsymbol{\Sigma}}$ and response modes $\tilde{\boldsymbol{Y}}$ by passing the forcing modes $\tilde{\boldsymbol{F}}$ to the full resolvent operator as $\tilde{\boldsymbol{Y}}\tilde{\boldsymbol{\Sigma}} = \boldsymbol{\mathcal{H}}_{\bar{\boldsymbol{q}}} \tilde{\boldsymbol{F}}$ for enhanced accuracy.

\begin{figure}
	\centering
    \includegraphics[width=1.0\textwidth, trim=0.162in 0.99in 0.162in 0.1in, clip]{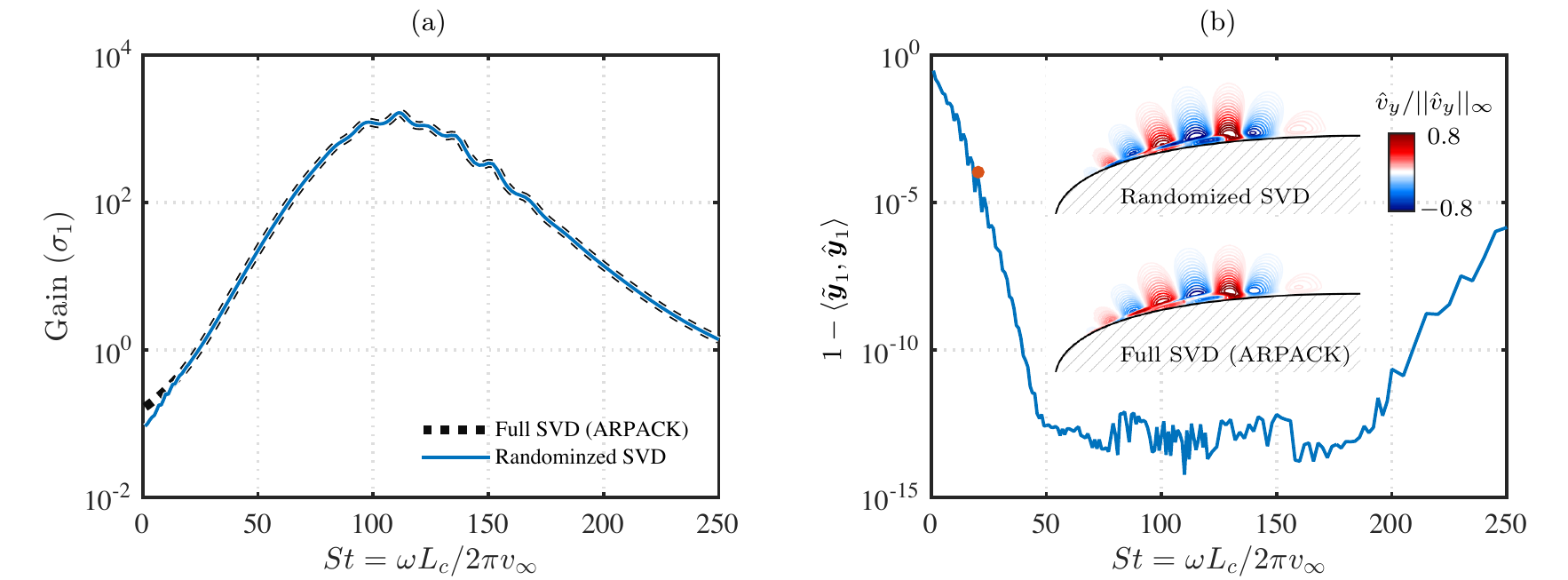}
	\caption{\label{fig:randSVD_demo} Comparison of the randomized SVD results to those from full SVD using ARPACK.  Agreements are found with respect to (a) gain profile and (b) cosine similarity between the response modes $\langle \tilde{\boldsymbol{y}}_1, \hat{\boldsymbol{y}}_1 \rangle$ over the range of examined frequency.  Here, we use $k_zL_c = 0$ and $\beta L_c/(2\pi v_\infty) = 20$, without spatial windowing, i.e. $\boldsymbol{B} = \boldsymbol{C} = \boldsymbol{I}$.  Inserted in (b) are the $\hat{v}_y$-components of the representative response mode at $St = 20$, highlighted with {\color{red2} $\bullet$} on the cosine similarity profile.}
\end{figure}

In this study, we use only $5$ test vectors ($k = 5$) to derive the low-rank approximation for $\boldsymbol{\mathcal{H}}_{\bar{\boldsymbol{q}}}$, achieving a significant compression of $ k / m \approx 5 \times 10^{-7}$.  We show the comparison of the results obtained from randomized SVD and full SVD (ARPACK package) in FIG.~\ref{fig:randSVD_demo}.  Here, we observe remarkable agreements with respect to the gain (singular value) profiles and resolvent modes (singular vectors) across a wide range of frequency.  The response modes at $St = 20$ are also inserted in FIG.~\ref{fig:randSVD_demo}b for comparison.  At an error level of $10^{-4}$, which is still far from the best accuracy achieved with the present randomized resolvent, we observe no visual discrepancy in the modal structure.  A slight difference in gain can be seen in $St \in [0, 10]$.  However, this frequency range is not of our primary interest, since the wavelength of the convective structure in this frequency range is at least twice the streamwise extent of the LSB, which is not observed over the LSB according to our flow simulation.  

With the randomized SVD, the computational cost and the memory requirement are significantly reduced.  These benefits allow us to conduct resolvent analysis for the present high-Reynolds number base flow in a computationally tractable manner.

\section{Results}
\subsection{Base flow: laminar separation bubble}

\begin{figure}
	\centering
    \includegraphics[width=1.0\textwidth, trim=0.162in 0.8in 0.162in 0, clip]{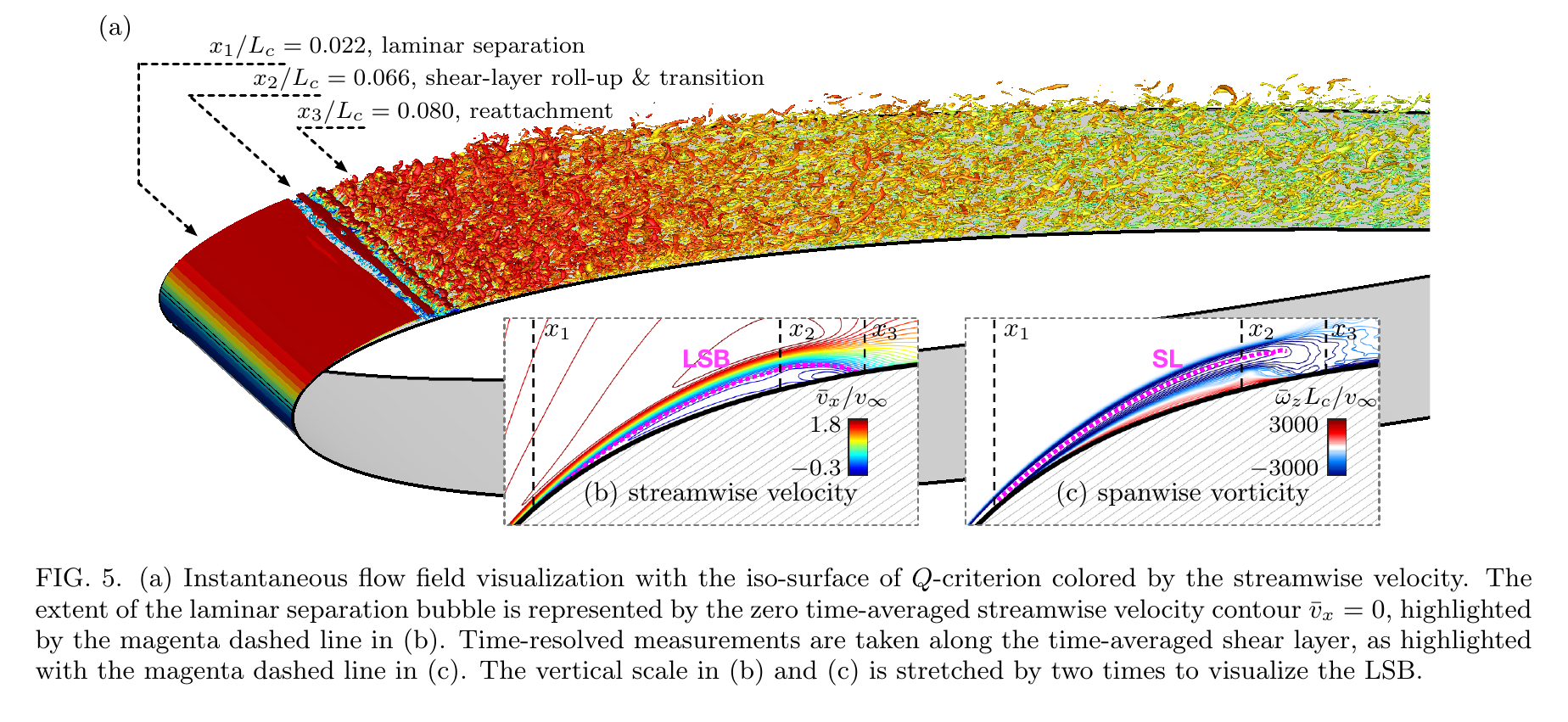}
	\caption{\label{fig:LSB_FlowVis} (a) Instantaneous flow field visualization with the iso-surface of $Q$-criterion colored by the streamwise velocity.  The extent of the laminar separation bubble is represented by the zero time-averaged streamwise velocity contour $\bar{v}_x = 0$, highlighted by the magenta dashed line in (b).  Time-resolved measurements are taken along the time-averaged shear layer, as highlighted with the magenta dashed line in (c).  The vertical scale in (b) and (c) is stretched by two times to visualize the LSB.}
\end{figure}

The focus of the current study is the laminar separation bubble (LSB) that forms over a NACA 0012 airfoil at $\al=8\dg$.  
The visualization of an instantaneous flow field using the iso-surface of $Q$-criterion is shown in FIG. \ref{fig:LSB_FlowVis}a.  
For $Re_{L_c}=500,000$ and $M_\infty=0.1$, the flow separates over the suction surface at $x_1/L_c = 0.022$.  
At the separation point, the displacement- and momentum-thickness-based Reynolds numbers are $Re_{\delta^*}=493$ and $Re_{\theta}=127$, respectively.  The flow separation forms a shear layer over the LSB.  At $x_2/L_c = 0.066$, the shear layer rolls up in the frequency range of $St \equiv \omega L_c/(2\pi v_\infty) \approx 100$.  At this location, the boundary layer reaches its maximum thickness over the LSB and transitions to turbulence.  This transition provides momentum mixing and reattaches the flow \citep{Marxen:JFM2011, Marxen:JFM2013, Yeh:JFM2019} at $x_3/L_c = 0.080$, giving the total streamwise extent of $0.058L_c$ to the LSB.  In the downstream of this streamwise location, the boundary layer remains attached and turbulent over the rest of the suction surface.  Meanwhile, the boundary layer on the pressure surface is laminar from the leading edge to the trailing edge. 

These three locations, $x_1$, $x_2$ and $x_3$, are marked by the gray dashed lines over the time-averaged streamwise velocity and spanwise vorticity fields in FIG. \ref{fig:LSB_FlowVis}b and \ref{fig:LSB_FlowVis}c.  Their respective flow features of separation, shear-layer roll-up, and reattachment can be clearly identified in FIG. \ref{fig:LSB_FlowVis}b by the magenta dashed line of zero time-averaged streamwise velocity.  In FIG. \ref{fig:LSB_FlowVis}c, we track the shear layer by the maximum spanwise vorticity and take probe measurements along this magenta line.  In the following discussions, we will revisit these three locations and the associated flow physics.  We will also compare the velocity spectra along the shear layer to the results of linear stability and resolvent analyses.

\subsection{Quasi-parallel linear stability analysis}
We perform the quasi-parallel linear stability analysis on the body-fitted coordinate system aligning with the surface-tangent ($\xi$) and surface-normal ($\eta$) directions, as shown in FIG. \ref{fig:Demo_LSA_vs_Resolvent}a.  The origin of the coordinate system is placed at the stagnation point, located slightly below the airfoil leading edge due to the positive $\al$ condition.  The analysis is conducted in each successive surface-tangent stations $\xi$ over the region where LSB resides, considering two-dimensional (2-D) perturbations ($k_zL_c=0$).

\begin{figure}
\vspace{0.0in}
\centering
    \includegraphics[width=1.0\textwidth, trim=0.162in 0.25in 0.162in 0.0in, clip]{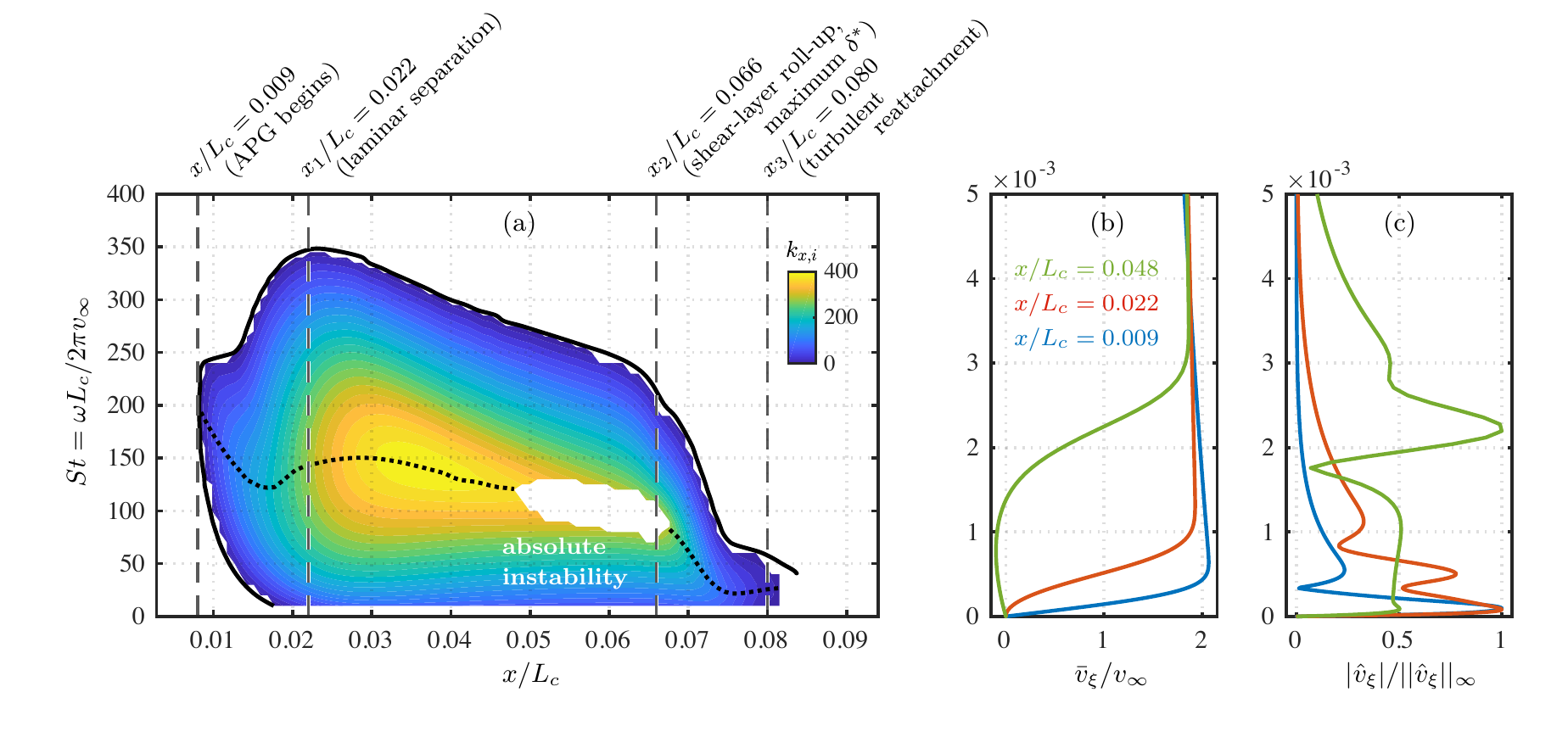}
	\caption{\label{fig:LSA_results} Results of quasi-parallel linear stability analysis. Shown are (a) the contour of local amplification rate $k_{x,i}$, (b) representative boundary layer profiles, and (c) the corresponding eigenfunctions. Eigenfunctions in (c) correspond to $St = 200$ at $x/L_c=0.009$ (first station of adverse pressure gradient) and $St = 120$ at $x/L_c = 0.022$ (separation point) and $0.048$ (upstream of absolute instability).  Dotted black line in (a) tracks the most amplified frequency at a given streamwise location.}
\end{figure}

The results of the linear stability analysis are shown in FIG. \ref{fig:LSA_results}a.  We find that instability first occurs at $x/L_c = 0.009$ near the frequency of $St = 200$.  This streamwise station corresponds to the point of peak suction and the beginning of the adverse pressure gradient (APG), which characterizes the boundary layer on the suction surface.  From this station to the laminar separation point $x_1/L_c=0.022$, the boundary layer remains laminar and attached, as suggested by the velocity profile in FIG. \ref{fig:LSA_results}b.  Correspondingly, we find that the instability in this region is of the Tollmien--Schlichting (T--S) type for an APG laminar boundary layer, which we recognize according to the eigenmode profile at $x/L_c = 0.009$ shown in FIG. \ref{fig:LSA_results}c.  The frequency associated with the maximum $k_{x,i}$ decreases to $St \approx 140$ at $x_1/L_c = 0.022$, reflecting the thickening of the boundary layer while marching downstream.

\begin{figure}
\centering
    \includegraphics[width=1.0\textwidth, trim=0.162in 0.6in 0.162in 0, clip]{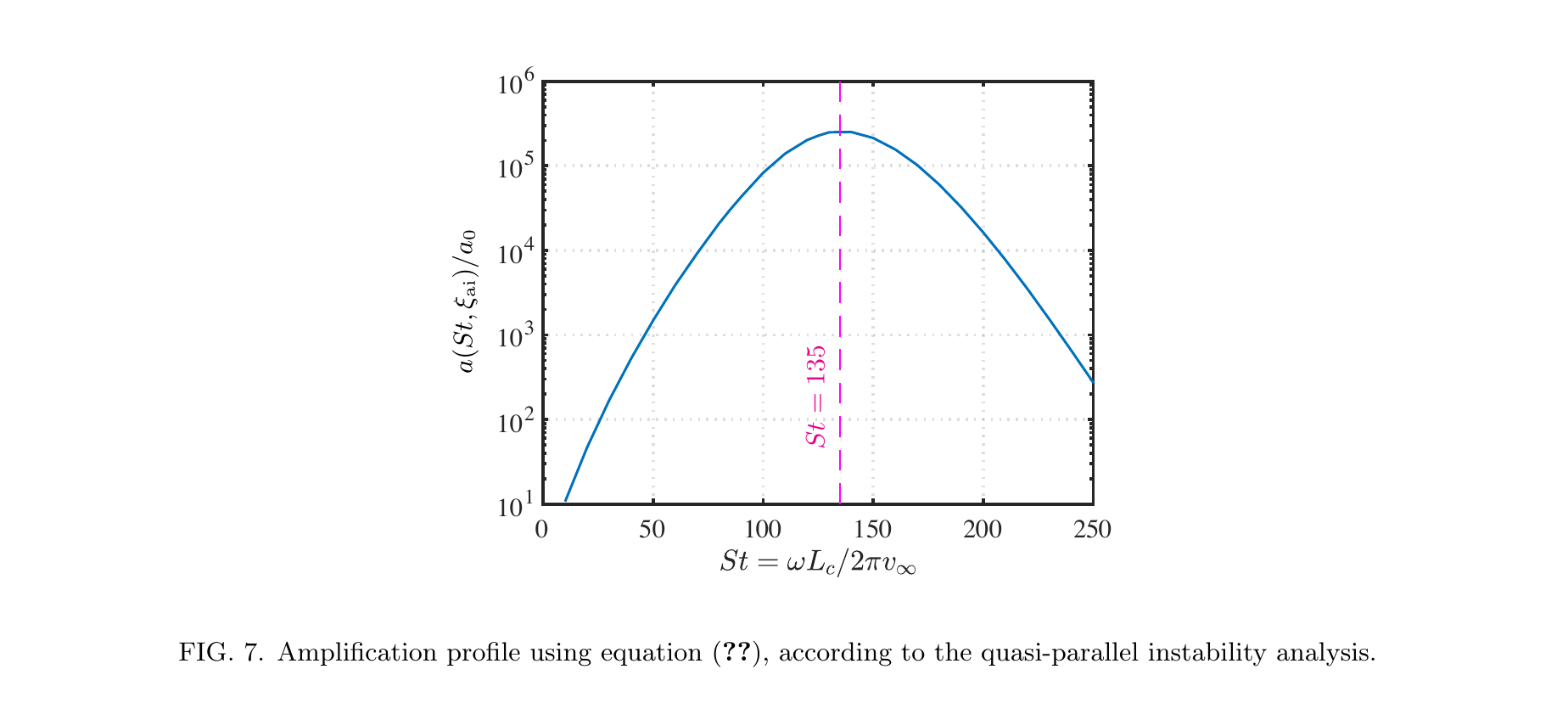}
	\caption{\label{fig:LSA_A0} Amplification profile using equation (\ref{eq:ampLST}), according to the quasi-parallel instability analysis. }
\end{figure}

From the laminar separation point ($x_1/L_c=0.022$) to the point of transition ($x_2/L_c=0.066$),  we observe the Kelvin--Helmholtz (K--H) type instability.  This region is characterized by significantly higher amplification rates and the development of reverse flow within the boundary layer.  The strong reverse flow over $x/L_c \in [0.05, 0.066]$ advects amplified perturbations back upstream, resulting in a closed region of absolute instability (amplification in both time and space) depicted by the white area in FIG. \ref{fig:LSA_results}a. We compute the total spatial amplification using equation (\ref{eq:ampLST}) for each frequency, up to the location $\xi_\text{ai}$ where absolute instability occurs.  This amplification profile, shown in FIG. \ref{fig:LSA_A0}, suggests that the amplification peaks at $St = 135$.   This peak frequency will be compared to the results of resolvent analysis and the shear-layer probe measurements shortly.

\subsection{Resolvent analysis}

The quasi-parallel stability analysis in the previous section showed both viscous and inviscid mechanisms as well as convective and absolute instabilities.  The complexity of this flow system motivates the use of global analysis that is capable of a more general description of these modal behaviors.  In this section, we investigate the dominant energy-amplification mechanism over the LSB using resolvent (input-output) analysis.  We will first perform the global stability analysis to show that $\boldsymbol{\mathcal{A}}_{\bar{\boldsymbol{q}}}$ is unstable and the discounting needs to be considered.  With the minimal discounting parameter obtained from the global stability analysis, we determine an appropriate range of the discounting parameter for the present unstable $\boldsymbol{\mathcal{A}}_{\bar{\boldsymbol{q}}}$.  Within this range, we will expand the response window over the LSB region and reveal that the dominant mechanism for energy-amplification is indeed the K--H instability along the shear layer. 

\subsubsection{Global stability analysis: minimal discounting}
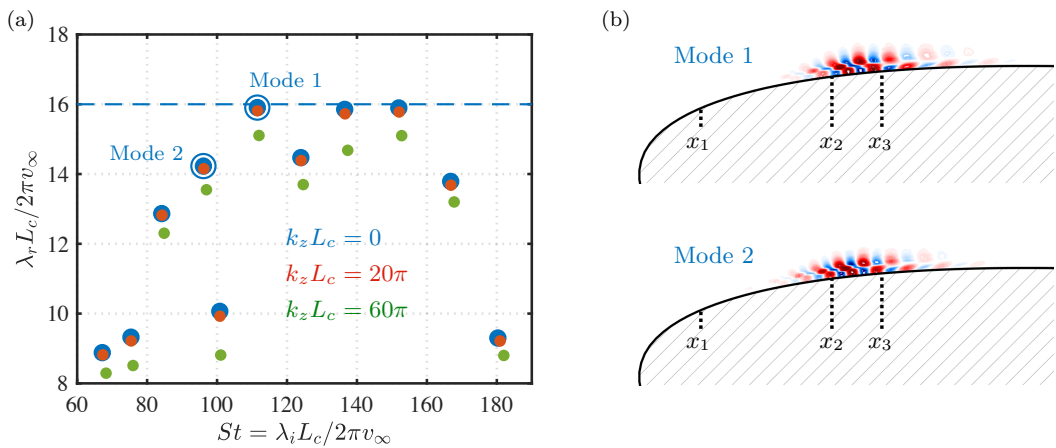
\begin{figure}
	\centering
	 \begin{overpic}[width=0.9\textwidth]{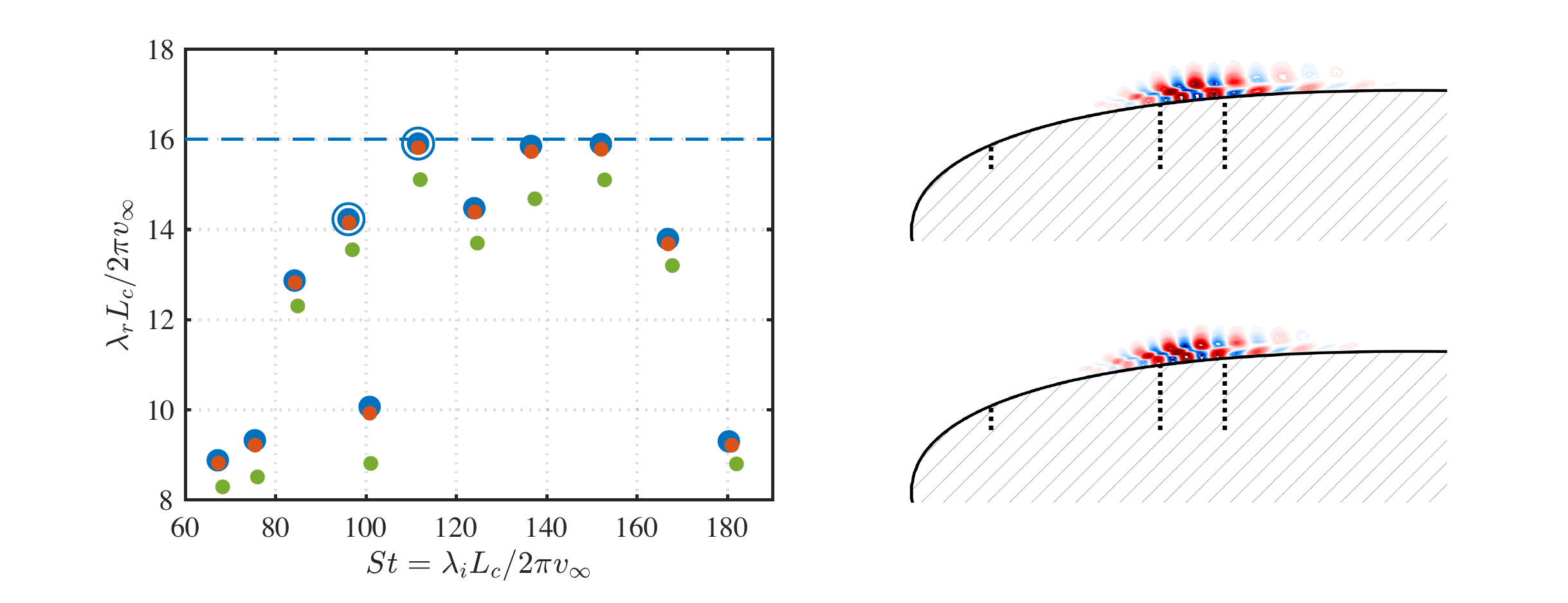}
		\put(06, 36){\footnotesize (a)}
		\put(29, 18){\color{blue2} $k_zL_c = 0$}
		\put(29, 15){\color{red2} $k_zL_c = 20\pi$}
		\put(29, 12){\color{green} $k_zL_c = 60\pi$}
		\put(26, 31){\footnotesize\color{blue2} Mode 1}
		\put(14.5, 25){\footnotesize\color{blue2} Mode 2}

		\put(55, 36){\footnotesize (b)}
		\put(61, 33){\color{blue2} Mode 1}
		\put(62, 26){$x_1$}
		\put(73, 26){$x_2$}
		\put(77, 26){$x_3$}
		
		\put(61, 16.5){\color{blue2} Mode 2}
		\put(62, 9.5){$x_1$}
		\put(73, 9.5){$x_2$}
		\put(77, 9.5){$x_3$}

	\end{overpic}
	\caption{\label{fig:GlobalModes} Global stability analysis for determining the minimal discounting parameter. (a) Eigenvalues of $\boldsymbol{\mathcal{A}}_{\bar{\boldsymbol{q}}}(k_z)$ appearing in the frequency range of LSB.  (b) Modal structures of two representative global modes ($\hat{\boldsymbol{v}}_x$-component, $k_z = 0$) circled in (a).}
\end{figure}

We perform the eigenvalue analysis of $\boldsymbol{\mathcal{A}}_{\bar{\boldsymbol{q}}}(k_z)$ to examine its stability characteristics as a precursor to the resolvent analysis to determine whether the discounting approach is needed.  This is equivalent to the global stability analysis of the mean flow $\bar{\boldsymbol{q}}$ at a given $k_z$.  The eigenvalues of $\boldsymbol{\mathcal{A}}_{\bar{\boldsymbol{q}}}(k_z)$ are shown in FIG. \ref{fig:GlobalModes}a for three selected spanwise wavenumbers of $k_z = 0$, $20\pi$ and $60\pi$.  We focus on the low wavenumber range since shear layers are known to be more unstable to long-spanwise-wave perturbations \citep{Pierrehumbert:JFM1982}. While the modal growth rate decreases with increasing $k_z$, we find that the linear operator $\boldsymbol{\mathcal{A}}_{\bar{\boldsymbol{q}}}(k_z)$ is unstable for all the considered $k_z$.  All the eigenmodes presented in FIG. \ref{fig:GlobalModes}a show fluctuations over the region of LSB, as represented by the two modal structures in FIG. \ref{fig:GlobalModes}b.  In particular, we find that both modes show the highest level of fluctuations over $x \in [x_2, x_3]$, suggesting that the dominant instability is associated with the shear-layer roll-up process.

Since the linear operator $\boldsymbol{\mathcal{A}}_{\bar{\boldsymbol{q}}}$ about the turbulent mean flow is found to be unstable, we consider the use of the discounting approach for the present resolvent analysis.  According to the spectra in FIG. \ref{fig:GlobalModes}a, the most unstable global mode of $\boldsymbol{\mathcal{A}}_{\bar{\boldsymbol{q}}}(k_z)$ appears at $\lambda L_c/(2\pi v_\infty) = 16.0+111i$ for $k_z = 0$.  This highest modal growth rate of $\lambda_r L_c/(2\pi v_\infty) = 16.0$ is determined as the minimal threshold for the discounting parameter.

\subsubsection{Discounting parameter}

With the minimal discounting determined, we start the analysis by seeking an appropriate range of the discounting parameter $\beta$.  Instead of quoting $\beta$, we report the results in an alternative form $t_\beta \equiv 2\pi / \beta$,  since this form of a time scale is more straightforward in providing insights into the convective structures of the resolvent modes.  As the largest modal growth rate is found to be $\lambda_r L_c/(2\pi v_\infty) = 16.0$, the longest allowable $t_\beta v_\infty / L_c = 0.0625$.  Therefore, we decrease $t_\beta$ from this point and characterize its effect by sweeping though a range of values for $t_\beta$.  

\begin{figure}
	\centering
    \includegraphics[width=1.0\textwidth, trim=0.162in 0.8in 0.162in 0.1in, clip]{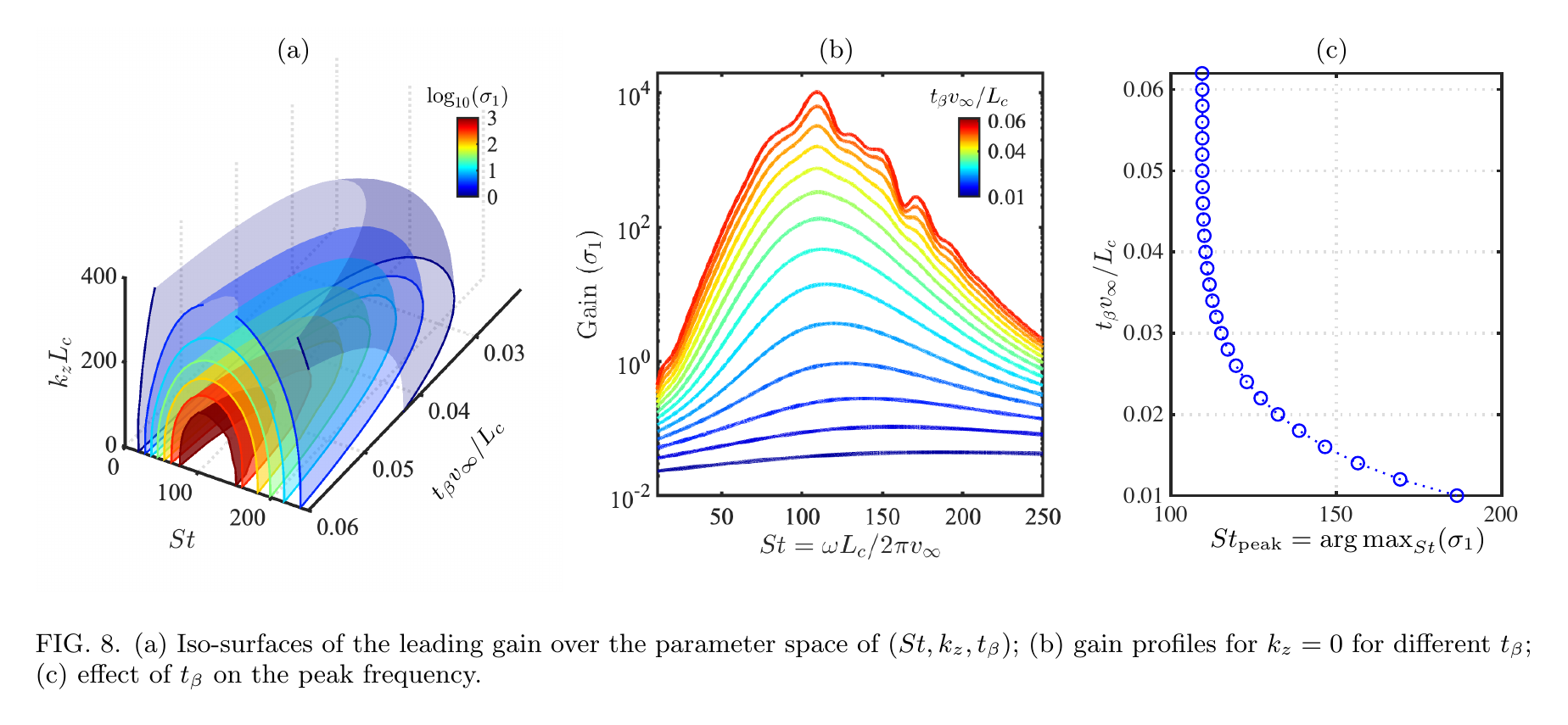}
	\caption{\label{fig:Gain_vs_TBeta} 
(a) Iso-surfaces of the leading gain over the parameter space of $(St, k_z, t_\beta)$; (b) gain profiles for $k_z = 0$ for different $t_\beta$; (c) effect of $t_\beta$ on the peak frequency.}
\end{figure}

\begin{figure}
\vspace{0.1in}
	\centering
    \includegraphics[width=1.0\textwidth, trim=0.162in 0.7in 0.162in 0.2in, clip]{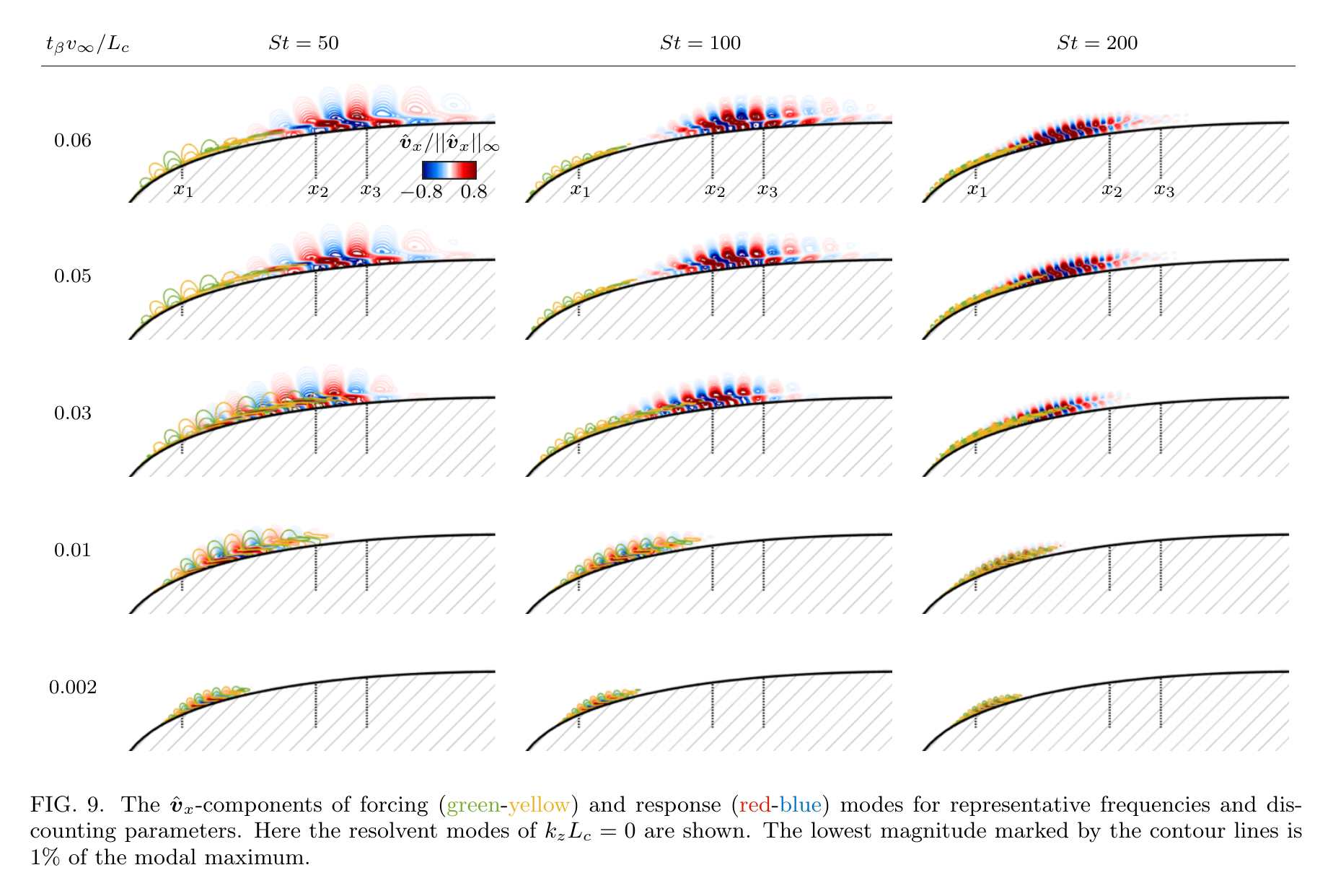}
	\caption{\label{fig:Modes_vs_TBeta} The $\hat{\boldsymbol{v}}_x$-components of forcing ({\color{green}green}-{\color{yellow}yellow}) and response ({\color{red2}red}-{\color{blue2}blue}) modes for representative frequencies and discounting parameters.  Here the resolvent modes of $k_zL_c = 0$ are shown. The lowest magnitude marked by the contour lines is $1\%$ of the modal maximum.}
\end{figure}

We sweep through the parameter space spanned by $(St, k_z, t_\beta)$ and present the gain profiles in FIG. \ref{fig:Gain_vs_TBeta}.   A general trend of decreasing leading gain ($\sigma_1$) with increasing $k_z$ and decreasing $t_\beta$ is observed in FIG. \ref{fig:Gain_vs_TBeta}a.  The trend of decreasing gain with the deceasing $t_\beta$ agrees with the previous studies \citep{Yeh:JFM2019,Sun:AIAAJ2019}, which can be explained by the exponentially growing amplification being evaluated within a shortening finite-time horizon due to the decreasing $t_\beta$.  We also find that the highest gain always appears at $k_z = 0$ for a given combination of $(St, t_\beta)$, agreeing with previous studies on shear layer instabilities \cite{Pierrehumbert:JFM1982,Yeh:JFM2019}.  As such, we turn our attention to the 2-D gain profiles in FIG. \ref{fig:Gain_vs_TBeta}b-c.  The 2-D gain profiles of $t_\beta v_\infty / L_c \in [0.01, 0.062]$ are shown in FIG. \ref{fig:Gain_vs_TBeta}b. At $t_\beta v_\infty / L_c = 0.062$, the gain profile has several extruding (yet smooth) peaks.  From the perspective of pseudospectrum, the appearance of these peaks suggests that this path of frequency sweeping lies in the proximity of the least stable eigenvalues.  At each $t_\beta$, we find the frequency at the highest leading gain and show the $t_\beta$-effect on the peak frequency in FIG. \ref{fig:Gain_vs_TBeta}c.  We observe that the peak frequency remains in the vicinity of $St = 110$ in $t_\beta v_\infty / L_c = [0.04, 0.062]$, suggesting an energy-amplification mechanism that is independent of the choice of $t_\beta$ in this range.

This mechanism can be revealed by studying the structures of the resolvent modes, as shown in FIG. \ref{fig:Modes_vs_TBeta}.  In the range of $t_\beta v_\infty / L_c \in [0.04, 0.062]$, we observe no apparent changes in the modal structure for all examined frequencies and wavenumbers, similar to the $t_\beta$-effect on the gain profiles in FIG. \ref{fig:Gain_vs_TBeta}b-c.   Within this range,  we find that the spatial supports for forcing and response modes are almost disjointed.  While the forcing modes cover the vicinity of the separation point $x_1$, the response modes exhibit high levels of fluctuations {\it downstream} of the forcing and extend their structures over the region of LSB.  This observation indicates that the dominant energy-amplification mechanism is the convective K--H instability which amplifies perturbations by the base flow advection.

When $t_\beta$ further decreases, the spatial supports of forcing and response gradually overlap.  At $t_\beta v_\infty /L_c = 0.02$ and beyond, the spatial support of the response mode lies on top of that of the forcing.   This overlap between forcing and response indicates that the energy amplification is dominated by the non-modal growth through the Orr mechanism for the short $t_\beta$.  Recently for jet flows, \citet{Schmidt:JFM2018} had offered similar discussions on the overlap between forcing and response and its relation with the Orr mechanism.  The tilting of the modal structure against the mean shear observed in the visualization is also a typical signature of the Orr mechanism \citep{Arratia:JFM2013,Garnaud:JFM2013,Schmidt:JFM2018}.

\citet{Arratia:JFM2013} concluded on the same dominant energy-amplification mechanisms for free shear layers over short and long time-horizons.  They examined the transient energy growth and showed that the amplification is dominated by the Orr mechanism over short-time horizons and by the K--H instability over long-time horizons.  While the resolvent analysis considers the harmonically forced problem instead of the initial value problem for transient growth, we find the interesting agreement here in the dominant mechanisms and the associated time horizons.  This agreement suggests that the discounting approach is capable of revealing the energy amplification mechanism in a forced system.  

We have noted that in FIG. \ref{fig:Gain_vs_TBeta}b the gain profiles in $t_\beta v_\infty/L_c \in [0.04, 0.062]$ peak near $St \approx 100$.  For short $t_\beta$ where the Orr mechanism becomes dominant, the peak frequency gradually moves to $St \approx 200$.  However, according to the ILES spectra, we find that the velocity spectrum along the shear layer exhibits peaks in the frequency range of $St \approx 100$.  This suggests that the long-time-horizon K--H mechanism dominates in the nonlinear flow.  In fact, the gain profiles in FIG. \ref{fig:Gain_vs_TBeta}b only show energy attenuation for $t_\beta v_\infty /L_c < 0.022$, which may explain the absence of high-frequency peaks that is associated with the Orr mechanisms in the nonlinear flow.  Moreover, the resolvent modes in FIG. \ref{fig:Modes_vs_TBeta} exhibit high levels of fluctuations between the roll-up location $x_2$ to reattachment point $x_3$ particularly for $St = 100$.  As being close to the peak-gain frequency in $t_\beta v_\infty / L_c \in [0.04, 0.062]$, this frequency has been shown to be effective in targeting the K--H instabilities of LSB to suppress the flow separation under dynamic stall in the previous flow control studies \citep{BentonVisbal:PRF2018, VisbalBenton:AIAAJ2018}.  Supported by these agreements, we conclude that the range of $t_\beta v_\infty / L_c \in [0.04, 0.062]$ is suitable for the present study on characterizing the energy-amplification mechanism over the LSB, as it properly reflects the effective K--H instability.   In what follows, we will focus on the use of $t_\beta v_\infty / L_c = 0.05$.

\subsubsection{Resolvent gain and velocity spectrum over the LSB}

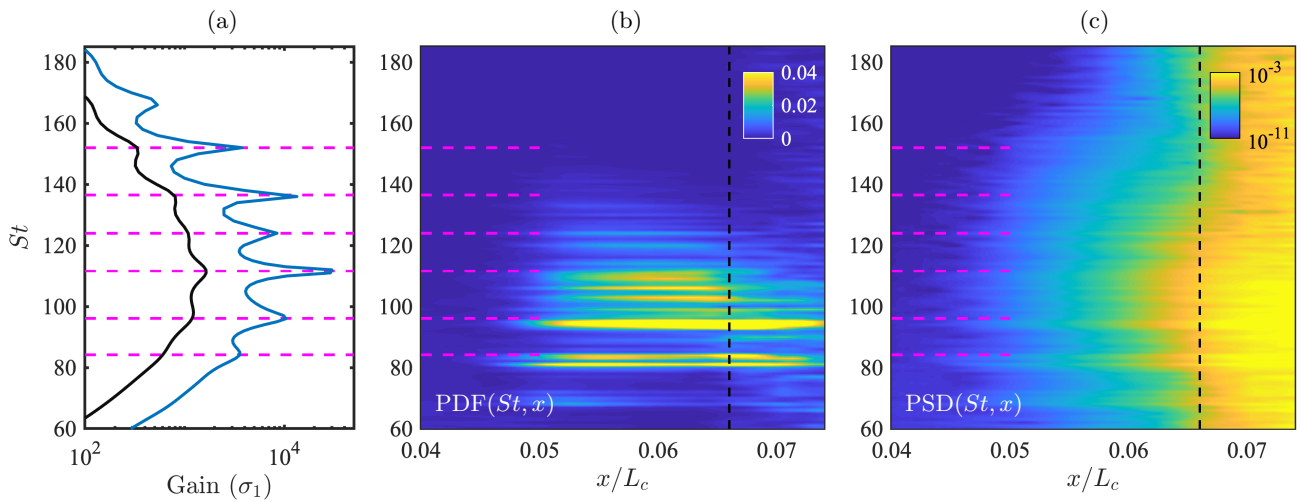
\begin{figure}
\begin{overpic}[scale=0.71]{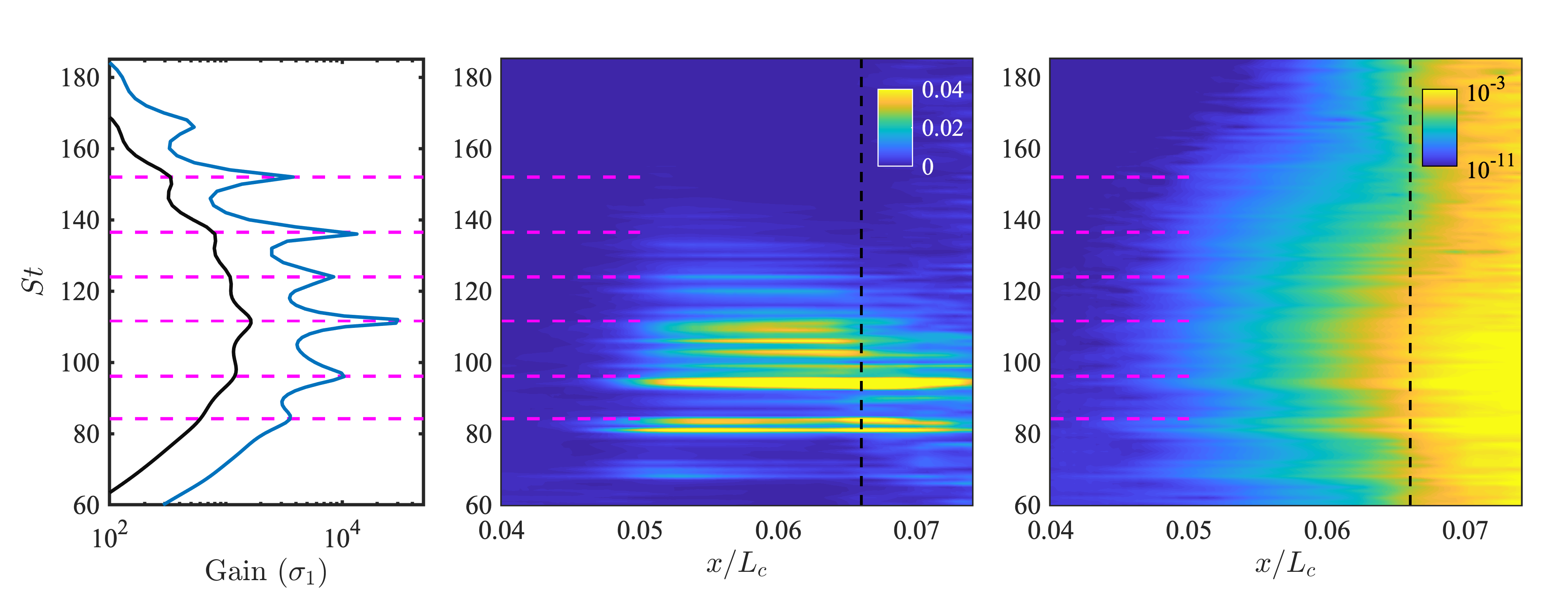}
	\put(16, 35.5){(a)}
	\put(46, 35.5){(b)}
	\put(81, 35.5){(c)}

	\put(33, 7){\color{white} $\text{PDF}(St, x)$}
	\put(68, 7){\color{white} $\text{PSD}(St, x)$}
\end{overpic}
\caption{\label{fig:GainPeak_LES} (a) Gain profiles of {$\color{black}t_\beta v_\infty / L_c = 0.05$} (black line) and $\beta = \max\left(\lambda_r(\boldsymbol{\mathcal{A}}_{\bar{\boldsymbol{q}}}))\right) + \epsilon$ (blue line), where $\epsilon = 10^{-4}$; (b) velocity spectra in terms of probability density function, and (c) in terms of power spectral density.}
\end{figure}

In addition to the gain profile obtained from $t_\beta v_\infty / L_c = 0.05$, shown in black in FIG. \ref{fig:GainPeak_LES}a, we also consider the use of the discounting parameter of $\beta = \max\left(\lambda_r(\boldsymbol{\mathcal{A}}_{\bar{\boldsymbol{q}}})\right) + \epsilon$, where $\epsilon = 10^{-4}$, to move the discounted frequency axis to the unstable eigenvalues of $\boldsymbol{\mathcal{A}}_{\bar{\boldsymbol{q}}}$ as close as possible.  The obtained gain profile is shown as the blue line in FIG. \ref{fig:GainPeak_LES}a.  We observe several sharp peaks in the gain profile appear with $\beta = \max\left(\lambda_r(\boldsymbol{\mathcal{A}}_{\bar{\boldsymbol{q}}}))\right) + \epsilon$.  From a pseudospectrum point-of-view, high level of gain will be achieved when evaluating the pseudospectrum in the vicinity of an eigenvalue, since the resolvent gain is unbounded at eigenvalues.  The gain profile with $t_\beta v_\infty / L_c = 0.05$, even farther away from the eigenvalues, signatures of these eigenvalues can still be observed by the slightly extruding gain profile at these frequencies. 


These peak frequencies obtained from resolvent analysis are compared to the velocity spectra from the probe measurements along the shear layer, as discussed in FIG. \ref{fig:LSB_FlowVis}b.  The spectra along the shear layer are shown with its probability density function (PDF) in FIG. \ref{fig:GainPeak_LES}b and power spectral density (PSD) in FIG. \ref{fig:GainPeak_LES}c in successive streamwise stations.  The laminar-turbulent transition at $x_2/L_c = 0.066$ can be identified in FIG. \ref{fig:GainPeak_LES}c, where the spectra becomes broadband beyond this streamwise station.  At the point of transition with broadband spectra, we still observe prominent spanwise coherent structure in the flow visualization in FIG. \ref{fig:LSB_FlowVis}a.  This is also reflected by the sharp peaks in the PDF in FIG. \ref{fig:LSB_FlowVis}b.  These peaks are compared to those obtained from the discounted resolvent analysis, marked by magenta dashed lines.  We find good agreement in these peak frequencies between the resolvent analysis and probe spectra.  The amplitude difference between the gain and probe data can be caused by the frequency-colored fluctuations sustained in the nonlinear flow \citep{Zare:JFM2017}, for which the resolvent analysis assumes white.  The presence of the additional peaks can be attributed to the subdominant modes observed in the global stability analysis and their nonlinear interactions with the dominant ones.  We also note that the peak frequency according to the quasi-parallel stability analysis is found to be $St = 135$ in FIG. \ref{fig:LSA_A0}.  However, this frequency component is not apparent in the probe measurement.  This comparison further values the use of the global resolvent approach to analyze the complex high-$Re$ flow.  

With the agreements between the resolvent gain and the velocity spectra over the LSB, we showed that the discounted analysis is capable of predicting the frequency content over the LSB.  Next, we compare the results to those from the non-discounted analysis and show that discounting is needed to land on these agreements.

\subsubsection{Discounting (temporal windowing) vs. spatial windowing}

Since the use of exponential discounting for unstable systems can be viewed as imposing a finite-time horizon in evaluating the input-output gain \citep{Jovanovic:Thesis2004}, its effect on resolvent analysis can also be interpreted as introducing a temporal filter, or window, that damps the output gain.  We have seen in FIG. \ref{fig:Modes_vs_TBeta} as well as from the previous study \citep{Yeh:JFM2019} that the response structure shrinks with the finite-time horizon, due to its convective nature.  While the restriction of the spatial extent on resolvent modes can also be achieved using spatial windowing in the input-output analysis, its effect on the resolvent analysis is not equivalent to that of temporal discounting.

\begin{figure}
	\centering
    \includegraphics[width=1.0\textwidth, trim=0.162in 0.65in 0.162in 0.15in, clip]{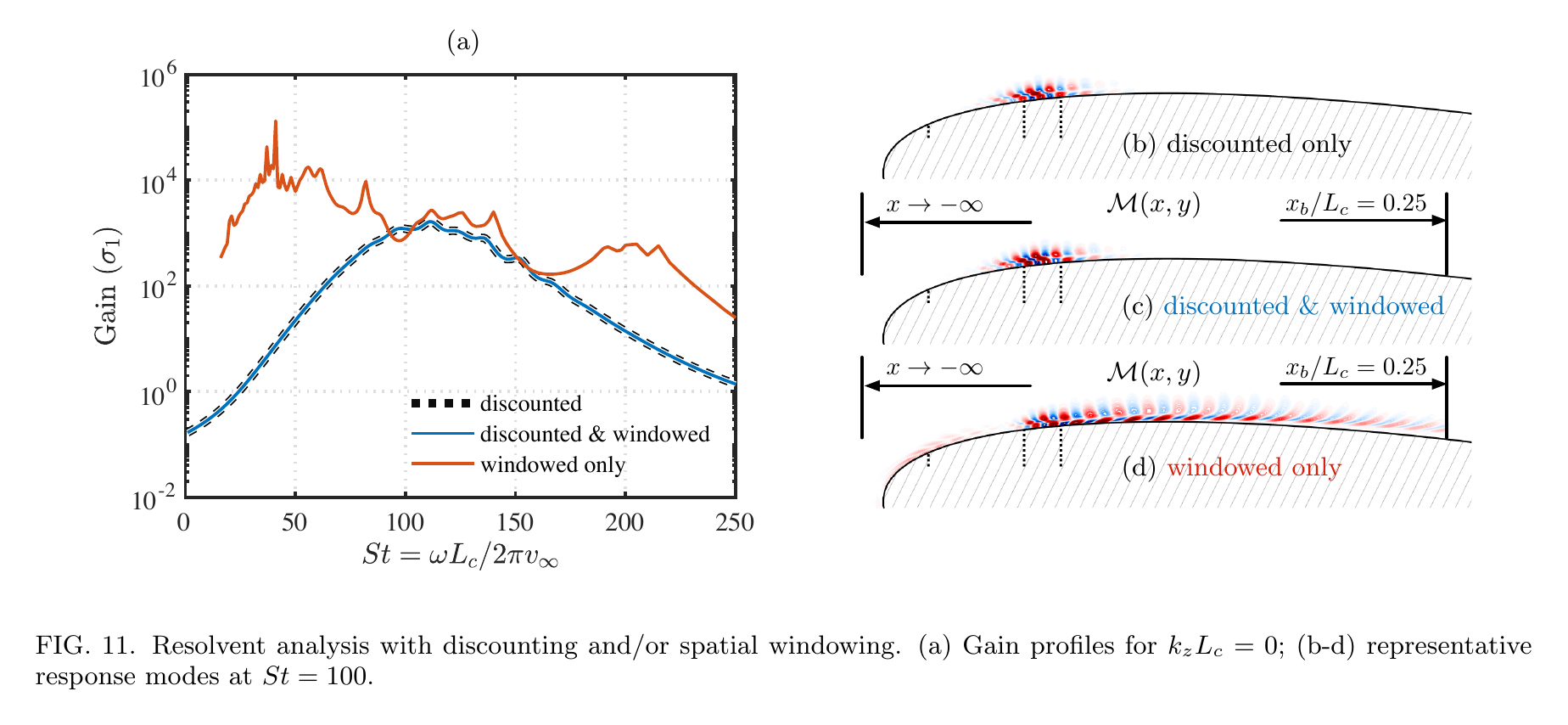}
	\caption{\label{fig:Discouting_vs_windowing} Resolvent analysis with discounting and/or spatial windowing. (a) Gain profiles for $k_zL_c = 0$; (b-d) representative response modes at $St = 100$.}
\end{figure}

In FIG. \ref{fig:Discouting_vs_windowing}, we show the gain profiles and the representative response modes at $St = 100$ obtained from the resolvent analyses using spatial windowing and/or temporal windowing (discounting).  Here, we apply the window only in the output matrix $\boldsymbol{C}$ due to the convective nature of the response while keeping the input matrix $\boldsymbol{B} = \boldsymbol{I}$.  The right boundary of the spatial window is chosen to be located at $x_b/L_c = 0.25$ and only the 2-D modes ($k_z = 0$) are examined.  We notice that the response mode obtained from the discounted analysis reside entirely in this spatial window, as shown in FIG. \ref{fig:Discouting_vs_windowing}b.  As such, the gain profile obtained from the use of the spatial window in addition to the same discounting perfectly collapses with that using discounting only, both revealing a highly amplified frequency range in the vicinity of $St \approx 100$ that is associated with the K--H instability over LSB.  We also observe no discrepancy in the response structure by comparing the response modes in FIG. \ref{fig:Discouting_vs_windowing}b to that in FIG. \ref{fig:Discouting_vs_windowing}c.

Applying only the spatial window does not provide similar results obtained from the other two.  Without discounting ($\beta = 0$), the resolvent analysis is conducted along the imaginary (frequency) axis.  From a pseudospectral point-of-view, the gain profile obtained along this path is influenced more by the subdominant eigenmodes than by the dominant ones that locate farther from this path on the complex plane.  Consequently, the gain profile in FIG. \ref{fig:Discouting_vs_windowing}a does not exhibit a similar trend over frequency, as it is influenced by subdominant modes populated close to the imaginary axis.  This gain profile struggles to reveal the dominant frequency range over the LSB in FIG. \ref{fig:GainPeak_LES}b, failing to provide accurate assessments on the energy-amplification characteristics of the LSB even with the applied spatial window. 

According to these findings and the agreements from the previous section, we showed that the appropriate use of discounting is needed in order to properly evaluate the energy-amplification characteristics of the LSB.  The discounting moves the frequency axis to the right of the least stable eigenmodes such that their dominance on the energy-amplification can be appropriately captured.  Such a procedure is also necessary considering the validity of the inverse Laplace transform on the steady-state output $\boldsymbol{y}$.  The discounting approach not only establishes a fair ground for assessing the energy amplification that takes place before it is overshadowed by the unbounded growth due to asymptotic instability, but also provides a zoom-in investigation of local physics from a global operator.  

\subsubsection{Output-windowed analysis}

\begin{figure}
	\centering
    \includegraphics[width=1.0\textwidth, trim=0.162in 0.65in 0.162in 0.1in, clip]{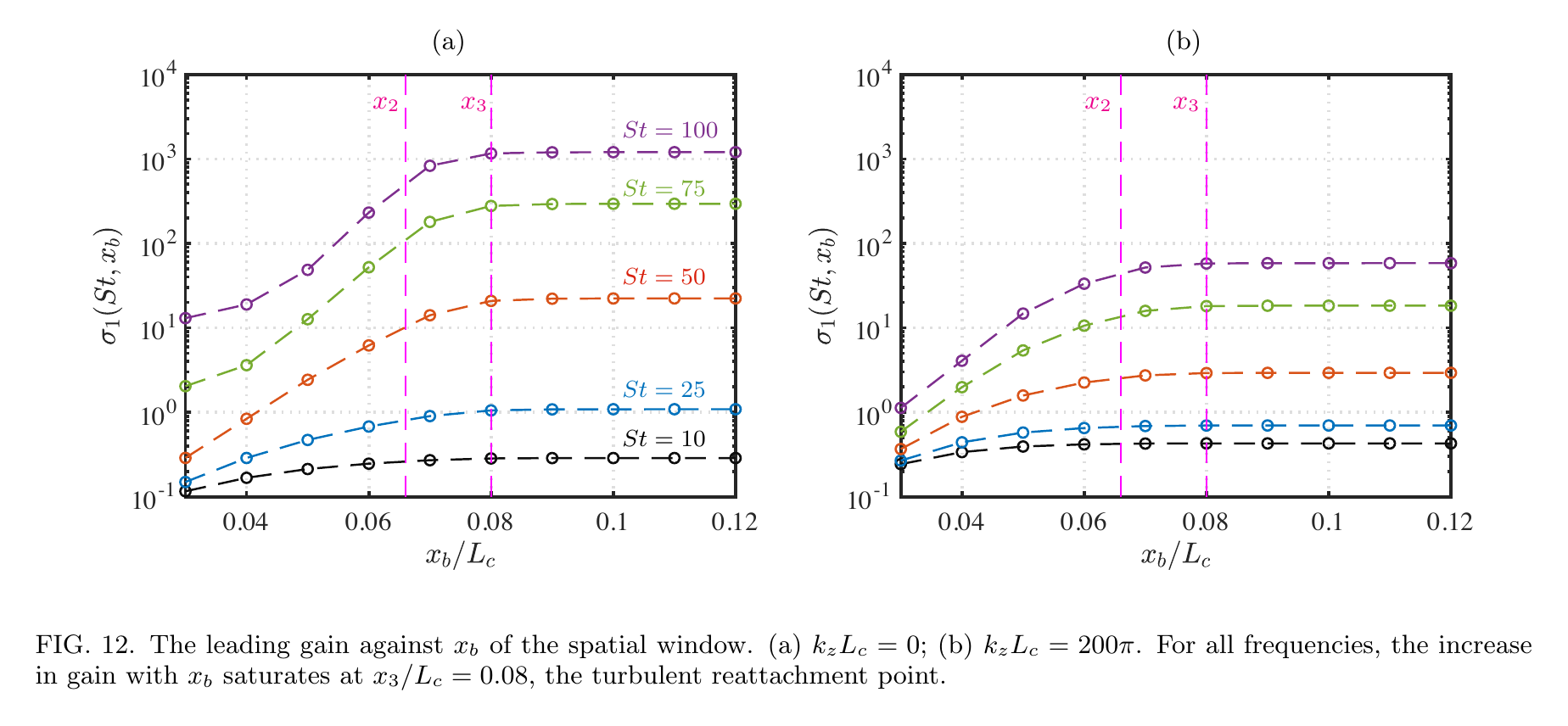}
	\caption{\label{fig:Gain_XB} The leading gain against $x_b$ of the spatial window. (a) $k_zL_c = 0$; (b) $k_zL_c = 200\pi$.  For all frequencies, the increase in gain with $x_b$ saturates at $x_3/L_c = 0.08$, the turbulent reattachment point. }
\end{figure}

In this section, we continue the use of the response window and keep $\boldsymbol{B} = \boldsymbol{I}$.  We fix the discounting parameter at $t_\beta v_\infty/L_c = 0.05$ and move the right-boundary of the spatial window, $x_b$, over the region of LSB from the separation point $x_1/L_c = 0.022$ to the turbulent reattachment point $x_3/L_c = 0.080$ and track how the energy-amplification changes with respect to $x_b$.
The results are show in FIG. \ref{fig:Gain_XB} with five representative frequencies and wavenumbers of $k_zL_c = 0$ and $200\pi$.  The gain for all frequencies continue to increase with increasing $x_b$, as driven by the convective K--H instability that allows perturbations to grow over the LSB while advecting downstream.  An important observation is that the increase in gain with $x_b$ saturates at the turbulent reattachment point $x_3/L_c = 0.080$.  While the results of $t_\beta v_\infty/L_c = 0.05$ are shown in FIG. \ref{fig:Gain_XB}, we observe no apparent increase in gain beyond the reattachment point of for all frequencies and wavenumbers in the range of $t_\beta v_\infty/L_c \in [0.04, 0.062]$.  This observation indicates that the dominant mechanism for energy-amplification over the LSB is indeed the K--H instability along the shear layer.  Once the flow transitions to turbulence and reattaches, there is no mechanism to support effective energy amplification over the suction surface.  This is an important consideration for flow control applications that target the flow instabilities to amplify the actuation input around the LSB.  The revealed mechanism suggests that the actuator needs to be placed upstream of the separation point to fully leverage the energy amplification over the LSB.

\begin{figure}
	\centering
    \includegraphics[width=1.0\textwidth, trim=0.162in 0.95in 0.162in 0.1in, clip]{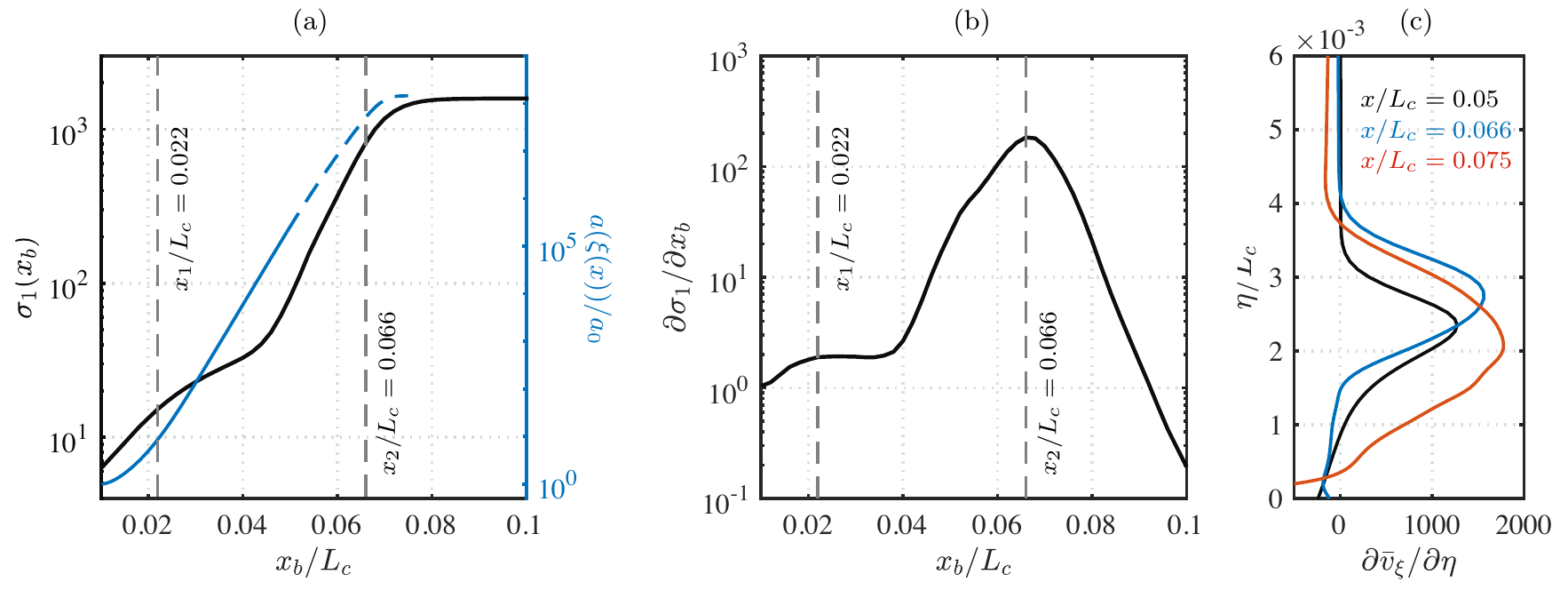}
	\caption{\label{fig:dGain_dXB} (a) The leading gain profile against the windowing $x_b$ for $St = 110$ and $k_zL_c = 0$. The energy amplification according to the linear stability analysis using equation (\ref{eq:ampLST}) is also shown for comparison, where the dashed part of the curve is obtained by interpolating for $k_{x, i}$ in the regions of absolute instability.  (b) The rate of change in the leading gain with respect to $x_b$.  (c) Base flow streamwise velocity profiles at three representative locations.}
\end{figure}

Besides the saturation of the increase in gain at $x_3/L_c = 0.080$, we also notice that the gain exhibits the steepest ramp-up at $x_2/L_c = 0.066$, where the shear layer rolls up and transition occurs.  This behavior is particularly prominent for frequencies of high gain, as seen for the case of $St = 100$ at $k_zL_c = 0$ in FIG. \ref{fig:Gain_XB}a.   In FIG. \ref{fig:dGain_dXB}, we focus on the case for $St = 110$ and $k_zL_c = 0$, which provides the highest gain over the spectral space spanned by $(St, k_z)$, and show the location of steepest gain increase.  Similar observation on the gain saturation at $x_3/L_c = 0.080$ can be made in FIG. \ref{fig:dGain_dXB}a for this frequency according to both resolvent analysis and linear stability analysis.  In particular, for this gain obtained from windowed resolvent analysis, we compute its spatial growth rate and show it in FIG. \ref{fig:dGain_dXB}b.  The spatial growth rate does exhibit the highest positive value at the shear-layer roll-up location, $x_2/L_c = 0.066$.  This is also suggested by the wall-normal gradient of local velocity profile at the same station, as shown in FIG. \ref{fig:dGain_dXB}c along with another two representative stations before and after the roll-up.  All of the velocity profiles show an inflection point to support the K--H instability.  At $x_2/L_c = 0.066$, the shear layer remains thin and compact, and the velocity gradient at the inflection point is higher than that of the upstream stations, represented by the profile at $x/L_c = 0.05$.  This high velocity gradient carried by the inflection point provides high spatial growth rate, according to invicid instability theory \citep{Monkewitz:PoF1982,HoHuerre:AR84}.  Although the inflection-point velocity gradient becomes even higher at $x/L_c = 0.075$, the thickened shear layer due to roll-up slows down the spatial growth rate, resulting in the steepest growth in gain at $x_2/L_c = 0.066$.

\subsubsection{Input-windowed analysis}

In this section, we apply the spatial window in the input matrix $\boldsymbol{B}$ while keeping the output matrix $\boldsymbol{C} = \boldsymbol{I}$ in order to study the optimal actuator location.  The input spatial window imposes the restriction that the forcing can only be introduced from the surface of the airfoil \citep{Garnaud:JFM2013}.  Moreover, we only introduce the forcing to the momentum equations, since we found that the forcing modes $\hat{\boldsymbol{f}} \equiv [\hat{\rho}_f, \hat{v}_{f_x}, \hat{v}_{f_y}, \hat{v}_{f_z}, \hat{T}_f] $ obtained from the previous sections typically have more than $95\%$ of the modal energy in their kinetic-energy components $[\hat{v}_{f_x}, \hat{v}_{f_y}, \hat{v}_{f_z}]$.  This form of forcing can be experimentally accomplished by various momentum-based actuators, including synthetic jets \citep{Glezer:ARFM2002,Cattafesta:ARFM11}.  The obtained forcing mode in this section will also provide insights
into the orientation of the jet-type momentum actuation.  

\begin{figure}
	\centering
    \includegraphics[width=1.0\textwidth, trim=0.162in 0.5in 0.162in 0.2in, clip]{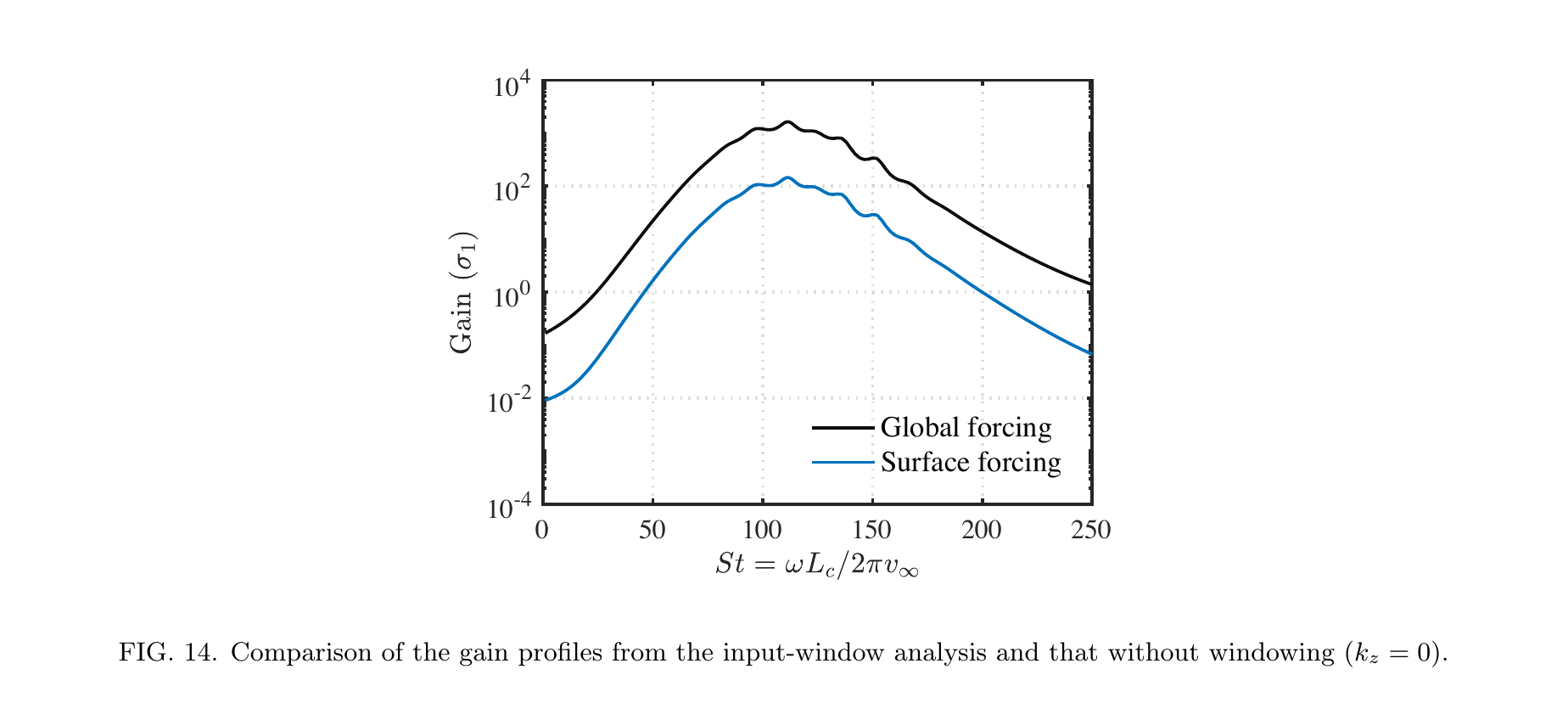}
	\caption{\label{fig:SurfF_Gain} Comparison of the gain profiles from the input-window analysis and that without windowing ($k_z = 0$).}
\end{figure}

The gain profile with the input window is shown as the blue line in FIG. \ref{fig:SurfF_Gain}.  Here, we keep our attention on the 2-D perturbations with $k_z = 0$.  Compared to the gain profile without windowing, the gain with the surface forcing is lowered approximately by an order of magnitude.  This is due to the fact that the forcing structure with the surface constraint is no longer the optimal forcing to the resolvent operator without the input window.  However, it is important that the gain profile conserves the overall trend over the entire frequency range of interest.

\begin{figure}
	\centering
    \includegraphics[width=1.0\textwidth, trim=0.162in 0.65in 0.162in 0, clip]{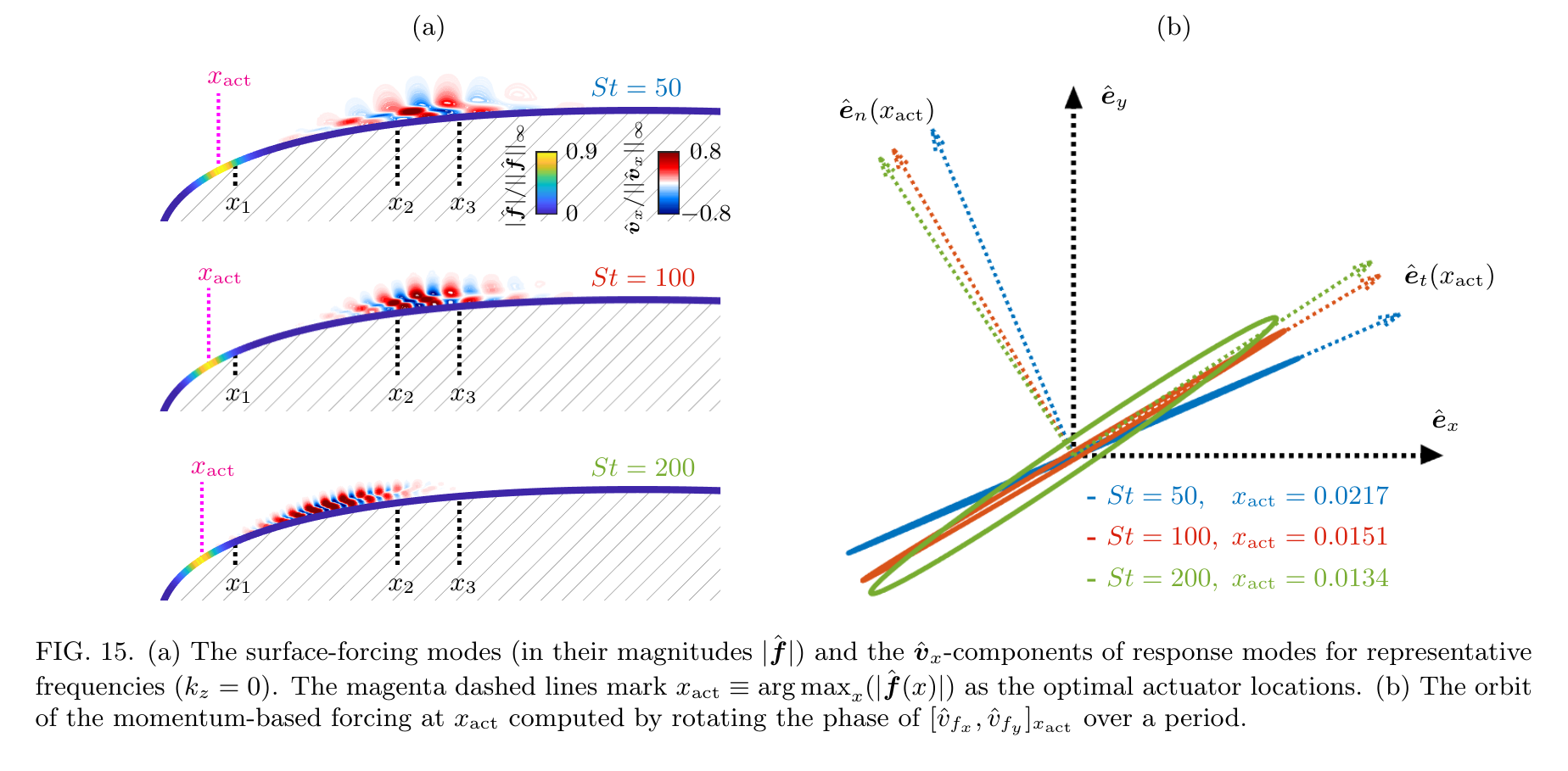}
	\caption{\label{fig:SurfF_Mode} (a) The surface-forcing modes (in their magnitudes $|\hat{\boldsymbol{f}}|$) and the $\hat{\boldsymbol{v}}_x$-components of response modes for representative frequencies ($k_z = 0$).  The magenta dashed lines mark $x_{\text{act}} \equiv \argmax_x (|\hat{\boldsymbol{f}}(x)|)$ as the optimal actuator locations. (b)  The orbit of the momentum-based forcing at $x_{\text{act}}$ computed by rotating the phase of $[\hat{v}_{f_x}, \hat{v}_{f_y}]_{x_{\text{act}}}$ over a period.}
\end{figure}

With the input window, the forcing mode structures provide interesting and important insights into the optimal actuator locations that render effective energy amplification, since they indicate sensitive regions in the flow.  These forcing modes are representatively shown in FIG. \ref{fig:SurfF_Mode}a by their magnitudes $|\hat{\boldsymbol{f}}|$ over the airfoil surface along with the corresponding response modes in the flow.  Compared those modes without windowing in FIG. \ref{fig:Modes_vs_TBeta}, the response modes do not exhibit observable changes.  For the forcing modes on the surface, we observe their highly compact spatial supports of approximately $1\%$ of the chord.  Moreover, by the magenta dashed line, we mark location $x_{\text{act}}$ where $|\hat{\boldsymbol{f}}|$ exhibits its highest magnitude, i.e. $x_{\text{act}} \equiv \argmax_x (|\hat{\boldsymbol{f}}(x)|)$, for each forcing mode.  As the most sensitive point of the flow, this point $x_{\text{act}}$ can be viewed as the optimal actuator location for the corresponding frequency.  We make an important observation here that $x_{\text{act}}$ is always upstream of the separation point.  This finding holds for the entire frequency range of interest and aligns with those from past studies \citep{Gad-el-Hak:JFE1991,Greenblatt:PAS2000,SeifertPack:AIAAJ1999,Alizard:PoF2009,BentonVisbal:PRF2018,Yeh:JFM2019}.  It is also physically intuitive in introducing perturbations at the onset of the shear layer.  In particular, we find that the optimal actuator locations $x_{\text{act}}$ only varies by less than $1\%$ of the chord length for $St \in [0, 250]$.  

We also examine the directionality of the momentum-based forcing modes at $x_{\text{act}}$ for the three representative frequencies.  The trajectories of these oscillatory forcings, shown as the solid orbits in FIG. \ref{fig:SurfF_Mode}b, can be computed by rotating the phase of $[\hat{v}_{f_x}, \hat{v}_{f_y}]_{x_{\text{act}}}$ over a period.  Along with these forcing orbits, the surface-normal vectors at the corresponding $x_{\text{act}}$ are also shown in the same color.  Here we make another important observation that the optimal forcing orientation is in the local {\it tangential} direction of the surface.  This observation agrees with the phase-reduction analysis by \citet{TairaNakao:JFM2018}, where they found that the surface tangent is the most effective forcing direction in synchronizing the vortex shedding in the cylinder wake with the periodic forcing introduced from the surface.  Moreover, by plotting these orbits in the same $\hat{\boldsymbol{e}}_x$-$\hat{\boldsymbol{e}}_y$ aspect ratio we find all of these orbits form ellipses of high eccentricity.   This is also an ideal feature for the momentum-based periodic forcing, since most of the actuators can only introduce uni-directional forcing 
\citep{Glezer:ARFM2002,Cattafesta:ARFM11}.  

In this section, we used the input window to impose the constraint that the forcing can only be introduced from the surface.  We showed that the optimal actuator location is upstream of the separation point, which enables the injection of perturbations at the onset of the shear layer.  Furthermore, we found that surface-tangent forcing is the optimal orientation for the momentum-based forcing.  The high uni-directionality of the forcing orbits also suggests a great potential for implementing active flow control with synthetic jets.  

\section{Conclusions}

We perform resolvent analysis to examine the energy-amplification characteristics over the laminar separation bubble (LSB) that forms over a NACA 0012 airfoil at a high $Re_{L_c} = 500,000$ and $\al = 8\dg$.  While we focus on the LSB residing over 6\% of the chord length, the resolvent operator is constructed about the global mean flow over the airfoil.  The resulting global resolvent operator for the high-$Re$ base flow has the big size of $10^6 \times 10^6$.  Therefore, we use the randomized SVD to perform resolvent analysis to relieve the computational cost.  With only $5$ sketching vectors, the randomized SVD provides high levels of accuracy in terms of gain and resolvent modes, promising its use in performing resolvent analyses on large-scale problems. 

To examine the local physics over the LSB, we consider the use of exponential discounting to limit the time horizon that allows for the instability to advect along the base flow.  In addition to discounting, we also examine the use of a spatial window that highlights the LSB in the resolvent analysis.  We find that the discounting can reveal the stability characteristics of the LSB due to the unstable base flow, regardless of the use of the spatial window.  With discounting, the gain distribution over frequency accurately captures the spectral content over the LSB obtained from the flow simulation. The peak-gain frequency also agrees with previous flow control studies on effectively suppressing dynamic stall over a pitching airfoil.  Supported by these findings, we show that discounting is capable of reveal local physics from a global operator and serves as an extension to resolvent analysis for unstable linear operators.

We also apply the spatial window to the response along with discounting in the resolvent analysis in order to reveal the energy-amplification mechanism over the LSB.  We move the downstream boundary of the spatial window across the extent of the LSB to allow for the convective instability to grow.  We find that the energy amplification grows over the LSB and saturates near the reattachment point.  The results show the dominant energy-amplification mechanism is the K--H instability over the LSB and agrees with physical insights from the base flow.  These findings confirm that the K--H instability dominate the energy amplification over LSB.  

Analogous to the response windowing, we also consider spatial windowing the input to impose the surface-forcing constraint for revealing the optimal actuator location.  In addition to this spatial constraint, the input window also confines the forcing only to the momentum equations.  We find that, according to the magnitude of the forcing mode, the optimal actuator location is always upstream of the separation point such that the forcing can be effectively introduced at the onset of the shear layer.  The forcing mode also sheds light on the optimal momentum forcing in the surface-tangent direction, with the high uni-directionality of forcing that is ideal for synthetic-jet-type actuators.  The physical insights provided by resolvent analysis can support the development of flow control strategies to target the LSB for suppressing flow separation and dynamic stall.

\begin{acknowledgments}

C.-A. Yeh and K. Taira thank the support from the Office of Naval Research (N00014-19-1-2460; Program managers: Dr.~D.~Gonzalez and Dr.~B.~Holm-Hansen) and the Air Force Office of Scientific Research (FA9550-18-1-0040 and FA9550-17-1-0222; Program managers: Dr.~G.~Abate and Dr.~D.~Smith).  C.-A. Yeh also acknowledges Profs.~Oliver Schmidt and Aaron Towne for their insightful feedbacks.  S. Benton and D. Garmann acknowledge support from the Air Force Office of Scientific Research under a Lab Task monitored by Dr. G. Abate.  Computational resources were provided by a grant of HPC time from the DoD HPC Shared Resource Centers at AFRL and ERDC.

\end{acknowledgments}

%

\end{document}